\documentclass{article}
\usepackage[utf8]{inputenc}%latin1
\usepackage[english]{babel}%english
\usepackage{dsfont}
\usepackage[cyr]{aeguill}
\usepackage{amsmath,amssymb}
\usepackage{amsthm}
\usepackage{eurosym}
\usepackage{multicol}
\usepackage{color}
\usepackage{fancyhdr}
\usepackage{graphicx}
\usepackage{verbatim}
\usepackage{caption}
\usepackage{stmaryrd}
\usepackage{float, multirow}
\usepackage[round]{natbib}
\usepackage[pdftex,colorlinks=automatic,linkcolor=black,citecolor=black,urlcolor=black]{hyperref}

\newcommand{\proba}{\mathbb{P}} %proba
\newcommand{\indic}{\mathds{1}} %indicatrice
\newcommand{\E}{\mathbb{E}} %esperance
\theoremstyle{plain}
\newtheorem{thm}{Theorem}
\newtheorem{pro}{Proposition}

\begin{document}

%%%%%%%%%%%%%%%%%%%%%%%%%%%%%%%%%%%%%%%%%%%% Title page
%\begin{titlepage}

\title{Nonparametric estimator of the tail dependence coefficient: \\ balancing bias and variance}

\author{
  Matthieu Garcin\thanks{Léonard de Vinci Pôle Universitaire, Research center, 92916 Paris La Défense, France,  \texttt{matthieu.garcin@m4x.org}  (corresponding author).}
  \and
  Maxime L. D. Nicolas \thanks{Université Paris I Panthéon-Sorbonne, Maison des Sciences Economiques, 106-112, boulevard de l'Hôpital 75013 Paris, France.}  
}

\date{\today}

\maketitle

\begin{abstract}
A theoretical expression is derived for the mean squared error of a nonparametric estimator of the tail dependence coefficient, depending on a threshold that defines which rank delimits the tails of a distribution. We propose a new method to optimally select this threshold. It combines the theoretical mean squared error of the estimator with a parametric estimation of the copula linking observations in the tails. Using simulations, we compare this semiparametric method with other approaches proposed in the literature, including the plateau-finding algorithm. 
\end{abstract}

%\noindent 

%\setcounter{page}{0}
%\thispagestyle{empty}
%\end{titlepage}

%%%%%%%%%%%%%%%%%%%%%%%%%%%%%%%%%%%%%%%%%%%% Introduction
\section{Introduction}

When considering several risk factors, risk managers across various fields such as finance, insurance, hydrology, and engineering, are interested in quantifying the dependence between all these random variables. Although the copula function is the most accurate description of dependence, a simple statistic is often preferred to this function in order to ease the interpretation. Popular examples of such a statistic include quantities based on a linear model, like the Pearson's correlation coefficient, or more realistic nonlinear approaches, such as Spearman's rho. However, these two examples do not specifically focus on the dependence between extreme events, and are thus not relevant in risk management applications. For instance, in finance, stronger dependencies between asset price returns are observed during recessions \citep{longin2001extreme,patton2004out}. Therefore, one might prefer using the tail dependence coefficient (TDC). The TDC depicts the probability that extreme events for several random variables happen simultaneously. It usually refers to the asymptotic probability introduced by \cite{sibuya1960bivariate} and later defined by \cite{joe1997multivariate}. It has been used for example in finance \citep{malevergne2003testing,poon2004extreme,CG} as well as in hydrology, for rainfall data \citep{poulin2007importance,serinaldi2008analysis,aghakouchak2010estimation}. It is worth noting that the purpose of TDC is not only to determine whether  data exhibit tail dependence or not, and hence what type of models might be suitable. There are indeed several applications which require a more accurate estimation of the TDC. For example, in finance, one can base the selection of a portfolio on the TDC, with the motivation of diversifying the portfolio with respect to extreme risks~\citep{de2011tail,de2017tail,WX}.

Before addressing the pivotal question of how to estimate the TDC, it is worth noting that the TDC is a pure copula property: it is not based on marginal distributions but only on the copula, that is the marginal-free version of the joint distribution \citep{nelsen2007introduction,joe2014dependence}. Therefore, the estimation of the TDC is strongly related to the estimation of the copula itself. 

The parametric estimation of the copula is a first solution that makes it possible to easily derive the TDC. The only challenging step is the choice of the copula function that best fits the data. Such a parametric procedure, in which the whole dataset is used to estimate the copula function, may not be appropriate since it does not focus on the tail. By exploiting extreme value theory, some parametric specifications of the copula however seem natural for depicting the dependence of extreme events \citep{einmahl2008method,kluppelberg2007estimating}. This is the case, for example, of the Clayton copula \citep{juri2002copula}.

To overcome the issue of choosing a specific parameterization of the copula function, some researchers proposed a nonparametric version of the TDC estimator based on the empirical copula introduced by \cite{Deheuvels}. This estimator corresponds to a discretization of the TDC as defined by \cite{joe1997multivariate} and relies on the selection of a threshold over which the probability of occurrence of joint extreme events is computed. \cite{coles1999dependence} have motivated a slightly different version of this nonparametric estimator, which is asymptotically equivalent. The selection rule for the threshold strongly impacts the quality of these nonparametric TDC estimators. Ideally, the threshold should make us focus on a few observations only, corresponding to extremes, in order to not bias the TDC estimation with data in the bulk of the distribution. However, the variance of the estimator would then be overriding. The threshold selection thus corresponds to the art of balancing adequately bias and variance. Most of the existing selection methods are heuristic. Among them, we can cite the plateau-finding algorithm \citep{FJS,SS2006}, or graphical methods \citep{CG}. Most of the contributions in the field are devoted to the comparison of various methods of TDC estimation in simulation frameworks \citep{FJS,SS2006,poulin2007importance,supper2020comparison}. Yet, to our knowledge, there is no theoretical contribution in which the selection rule of the threshold is related to a simple trade-off between the bias and the variance of the estimator. 

We thus propose a theoretical expression for both the bias and the variance of the nonparametric TDC estimator. We then use these expressions to define selection rules in which the threshold in the nonparametric TDC estimator minimizes the theoretical mean squared error (MSE). The formulas depend on the true and unobserved copula. Therefore, a practical application requires choosing a parametric specification for the copula, but only for the tails of the multivariate distribution. To this end, we consider two widespread Archimedean copulas, the Clayton and Gumbel copulas. The Clayton copula offers a flexible representation of tail dependence with various degrees of intensity. \cite{SS2006} proposed the Clayton copula to model the tail copula function. \cite{juri2002copula,juri2003tail} showed that the survival Clayton copula is a natural limit for joint excesses beyond a threshold having an Archimedean copula dependence structure. The Gumbel copula is also a natural choice to model upper tail dependence \citep{galambos1978asymptotic,joe1997multivariate} since it is the only copula that is at the same time Archimedean and an extreme-value copula \citep{genest1989characterization}. 

The paper is organized as follows. Section 2 rapidly recalls some basic definitions of the TDC. Section 3 is devoted to theoretical expressions for the bias and variance of nonparametric TDC estimators. Section 4 explores several selection rules for the threshold in the nonparametric TDC estimator. In the simulation study of Section 5, the performance of these estimators is shown to be similar to the one of the plateau-finding algorithm. Section 6 presents a short empirical application to financial data. Section 7 concludes.

\section{Tail-dependence coefficient}
%\subsection{Copula and Tail Dependence}

\cite{sklar1959fonctions} showed that any joint distribution of the pair $(X,Y)$ of real random variables can be written as a function of continuous marginal distributions:

$$ F(x,y)=C\left(F_X(x), F_Y(y)\right),$$

where $C$ is the copula function between $X$ and $Y$ and can be expressed as:
$$C(u,v) = F\left(F_X^{-1}(u), F_Y^{-1}(v)\right),$$
where $(u,v)\in\left[0, 1\right]^2$, $ F_X^{-1}$ and  $F_Y^{-1}$ are the inverse of the univariate distribution functions $F_X$ and $F_Y$. The dependence structure is fully described by the copula function and holds independently of the marginal distributions.

%\subsection{Tail dependence}

In a pioneering article, \cite{sibuya1960bivariate} introduced the notion of tail dependence. This notion describes the dependence between extreme values, either in the upper-right-quadrant tail or in the lower-left-quadrant tail of a bivariate distribution. The lower TDC, denoted $\lambda_L$, is defined as follows \citep{joe1997multivariate}:
$$\lambda_{L}=\lim _{u \rightarrow 0^{+}} \mathbb{P}\left[X<F_{X}^{-1}(u) | Y<F_{Y}^{-1}(u)\right]=\lim _{u \rightarrow 0^{+}} \frac{C(u, u)}{u},$$
if the limit exists. Similarly, the upper TDC is defined by:
$$\lambda_{U}=\lim _{u \rightarrow 1^{^{-}}} \mathbb{P}\left[X>F_{X}^{-1}(u) | Y>F_{Y}^{-1}(u)\right] =\lim _{u \rightarrow 1^{^{-}}} \frac{1-2 u+C(u, u)}{1-u},$$
if the limit exists. Since $\lambda_L$ and $\lambda_U$ are probabilities, they belong to $[0,1]$.

The tail dependence is a pure copula property, that is, it is independent of the margins of $X$ and $Y$. The TDC exists if the limits in the above equations exist. If $\lambda_L>0$ (respectively $\lambda_U>0$), then the copula presents tail dependence and we simultaneously observe extremely small (resp. extremely large) realizations of $X$ and $Y$, with conditional probability $\lambda_L$ (resp. $\lambda_U$). In contrast, the absence of tail dependence corresponds to the TDC equal to zero. In this case, the variables $X$ and $Y$ are asymptotically independent. 

\section{Nonparametric estimation of the TDC}

The estimation of the TDC is often related to the estimation of copulas. Indeed, if one estimates a parametric copula, one can easily deduce the corresponding parametric TDC. Nonetheless, an accurate estimation of the TDC requires focusing merely on extreme observations\footnote{ Otherwise, if one considers that all the observations, even not extreme, are linked by the same copula, other nonparametric methods including all the data are possible. For instance, one may use a nonparametric estimator of Pickands' dependence function, which is an important feature in the definition of an extreme-value copula \citep{caperaa1997,FJS}.}. However, the more one confines oneself to extreme observations, the less robust the estimator. The need to rely on extreme observations must thus be balanced with the equally important need to use a sufficiently large amount of data. Nonparametric techniques seem suitable for this purpose, as emphasized by \cite{joe1992bivariate}. In this section, we present nonparametric estimators and we introduce their corresponding MSE. We first focus on the lower TDC, then on the upper TDC, and we finish with an extension in which the estimator is itself an average of nonparametric estimators.

\subsection{Lower tail}

We are given $n$ bivariate observations $(X_j,Y_j)$, for $j\in\llbracket 1,n\rrbracket$, generated with a dependence model of copula $C$. The nonparametric estimator of the lower TDC is defined by the following~\citep{CG,FJS}:
\begin{equation}\label{eq:NonParmLambdaL}
\widehat \lambda_{L,n}\left(\frac{i}{n}\right)=\frac{\widehat C_n\left(\frac{i}{n},\frac{i}{n}\right)}{\frac{i}{n}},
\end{equation}
where $(u,v)\in[0,1]^2\mapsto\widehat C_n\left(u,v\right)$ is the empirical copula, introduced by \cite{Deheuvels}. We can write this empirical copula as follows \citep{GR}:
$$\widehat C_n\left(u,v\right)=\frac{1}{n}\sum_{j=1}^n{\indic\{\widehat F_{X,n}(X_j)\leq u\} \indic\{\widehat F_{Y,n}(Y_j)\leq v\}},$$
where $\widehat F_{X,n}$ and $\widehat F_{Y,n}$ are estimations of the marginal cumulative distribution functions. Focusing on $X$, $\widehat F_{X,n}$ is defined by:
$$\widehat F_{X,n}(x) = \frac{1}{n}\sum_{j=1}^n{\indic\{X_{j}\leq x\}}.$$
\cite{SS2006} have shown that the nonparametric estimator of the TDC has a strong consistency and is asymptotically normal. 

The estimator of the lower TDC provided in equation~\eqref{eq:NonParmLambdaL} relies on the selection of an appropriate integer $i\in\llbracket 1,n\rrbracket$. Various selection rules for this free parameter have been proposed in the literature. We can cite for example the plateau-finding algorithm \citep{FJS} or a graphical method based on monotonic variations of the estimator \citep{CG}.

Before depicting a new selection criterion for $i$ in equation~\eqref{eq:NonParmLambdaL}, we have to specify the role of this free parameter in the estimator. The definition of the lower TDC corresponds to the limit case $i/n\rightarrow 0$. Nevertheless, as exposed above, using the lowest possible value for $i$ would lead to a non-robust estimator. In contrast, a higher value of $i$ would depict some properties of the copula which are not specifically the ones of its lower tail. Therefore, these two effects, variance and bias, should be balanced in a good compromise. We thus intend to minimize the MSE between the estimator $\widehat \lambda_{L,n}\left(i/n\right)$ and the true lower tail dependence parameter $\lambda_L$. The value of this error is provided in Theorem~\ref{th:ErrorEstim} in an asymptotic framework.

We introduce some notation that will be used in the theorem. The diagonal section of the copula $C$ is $u\mapsto \delta(u)=C(u,u)$, with the corresponding nonparametric estimator $\widehat{\delta}_n$. The h-function of the copula is the conditional cumulative distribution function provided by $h_1(u,v)=\partial C(u,v)/\partial u$ and $h_2(u,v)=\partial C(u,v)/\partial v$. In the case of a symmetric copula, we will simply write $h(u,v)=h_1(u,v)=h_2(u,v)$. Moreover, the diagonal version of these h-functions, that is when $u=v$, is simplified in $h_1(u)$, $h_2(u)$, and $h(u)$.

\begin{thm}\label{th:ErrorEstim}
Let the bivariate copula $C$ have continuous partial derivatives and $i(n)$ be equal to $\alpha n$, where $\alpha\in(0,1)$. Then, the MSE of the nonparametric estimator of the lower TDC, defined in equation~\eqref{eq:NonParmLambdaL}, behaves asymptotically in the following manner:
$$\E\left[\left(\widehat \lambda_{L,n}\left(\frac{i(n)}{n}\right)-\lambda_L\right)^2\right] = V_{L,n}(\alpha) + \left(\frac{1}{\alpha}\delta\left(\alpha\right)-\delta'(0)\right)^2,$$
where
$$\underset{n\rightarrow\infty}{\lim} nV_{L,n}(\alpha)=\frac{\sigma^2\left(\alpha\right)}{\alpha^2}$$
and
\begin{equation}\label{eq:sigmaL}
\sigma^2(\alpha)=\delta(\alpha)(1-\delta(\alpha)) +(1-\alpha)\left[\alpha\left(h_1(\alpha)^2+h_2(\alpha)^2\right)- 2 \delta(\alpha)(h_1(\alpha)+h_2(\alpha))\right] + 2 h_1(\alpha)h_2(\alpha)(\delta(\alpha)-\alpha^2).
\end{equation}
\end{thm}

The proof is postponed to Appendix~\ref{sec:proofErrorEstim}. In the proof, a technical condition is required, specifically regarding the uniform integrability of $n\left(\widehat{\delta}_{n}(i(n)/n)-\delta(i(n)/n)\right)^2$, which is generally fulfilled as exposed in Appendix~\ref{sec:unifint}.

In Theorem~\ref{th:ErrorEstim}, the variance of the estimator is $\sigma^2\left(\alpha\right)/n\alpha^2$ and the squared bias is $\left(\delta\left(\alpha\right)/\alpha-\delta'(0)\right)^2$. This means that, given $\alpha\in(0,1)$, the variance will shrink to zero as $n\rightarrow\infty$ but not the bias. The squared bias and the variance will be more balanced for values of $\alpha$ close to zero and datasets of finite size $n$, for which we will apply this asymptotic framework. If the copula $C$ is symmetric, $\sigma^2(u)$ more easily writes:
\begin{equation}\label{eq:SymVar}
\sigma^2(u)=\delta(u)(1-\delta(u)) +2(1-u)h(u)\left[uh(u)-2 \delta(u)\right] + 2 h(u)^2(\delta(u)-u^2).
\end{equation}

We now apply Theorem~\ref{th:ErrorEstim} to the Clayton copula, which is symmetric:
$$C(u,v)=\left(u^{-\theta}+v^{-\theta}-1\right)^{-1/\theta},$$
where $\theta>0$. 

\begin{pro}\label{pro:Clayton}
In the case of the Clayton copula of parameter $\theta>0$, the asymptotic variance and squared bias of the nonparametric estimator $\widehat \lambda_{L,n}\left(i(n)/n\right)$ of the lower TDC, defined in equation~\eqref{eq:NonParmLambdaL}, with the assumptions of Theorem~\ref{th:ErrorEstim}, are:
$$\left\{\begin{array}{ccl}
\text{variance} & = & \frac{1}{n\alpha^2} \left(\delta(\alpha) - \delta(\alpha)^2\left[1+2\frac{2(1-\alpha)(1-\alpha^{\theta})+1}{\alpha(2-\alpha^{\theta})^2}\right] + 2\delta(\alpha)^3\left[\frac{1}{\alpha^2(2-\alpha^{\theta})^2}\right]\right) \\
\text{bias}^2 & = & \left(\left(2-\alpha^{\theta}\right)^{-1/\theta}-2^{-1/\theta}\right)^2,
\end{array}\right.$$
where $\delta(\alpha)=\left(2\alpha^{-\theta}-1\right)^{-1/\theta}$.
\end{pro}

The proof is postponed to Appendix~\ref{sec:proofClayton}.

\subsection{Upper tail}

We can extend to the upper tail the results exposed above for the lower tail. Starting from the relation between the survival copula $\bar C$ and the copula $C$,
$$\bar C(u,v)=u+v-1+C(1-u,1-v),$$
we note that the survival diagonal section is
$$\bar{\delta}(u)=2u-1+\delta(1-u).$$
The upper TDC is the lower TDC of the survival copula \citep{SS2006}. So we have
$$\lambda_U=\bar{\delta}'(0)=2-\delta'(1).$$
Remarking that
$$\delta'(1)=\underset{t\rightarrow 1^{^-}}{\lim}\frac{1- C\left(t,t\right)}{1-t}$$ 
the estimator of the upper TDC naturally follows:
\begin{equation}\label{eq:NonParmLambdaU}
\widehat{\lambda}_{U,n}\left(\frac{i}{n}\right)=\frac{1-2 \frac{i}{n}+\widehat{C}_n\left(\frac{i}{n}, \frac{i}{n}\right)}{1-\frac{i}{n}},
\end{equation}
which also depends on the selection of an appropriate $i\in\llbracket 1,n\rrbracket$.

\begin{thm}\label{th:ErrorEstimU}
Let the bivariate copula $C$ have continuous partial derivatives and $i(n)$ be equal to $\alpha n$, where $\alpha\in(0,1)$. Then, the MSE of the nonparametric estimator of the upper TDC, defined in equation~\eqref{eq:NonParmLambdaU}, behaves asymptotically in the following manner:
$$\E\left[\left(\widehat \lambda_{U,n}\left(\frac{i(n)}{n}\right)-\lambda_U\right)^2\right]  = V_{U,n}(\alpha)+\left(\frac{1-2 \alpha+\delta\left(\alpha\right)}{1-\alpha} - 2+\delta^{\prime}(1) \right)^{2},$$
where 
$$\underset{n\rightarrow\infty}{\lim} nV_{U,n}(\alpha)=\frac{\sigma^{2}\left(\alpha\right)}{(1-\alpha)^{2}}$$
and $\sigma^2(\alpha)$ is the same as in Theorem~\ref{th:ErrorEstim}.
\end{thm}

The proof is postponed to Appendix~\ref{sec:proofErrorEstimU}.

Like for Theorem~\ref{th:ErrorEstim}, we can explicitly split the MSE of the nonparametric estimator of the upper TDC, expressed in Theorem~\ref{th:ErrorEstimU}, in two components: the variance $\sigma^{2}\left(\alpha\right)/n(1-\alpha)^{2}$ and the squared bias $\left(\left(1-2 \alpha+\delta\left(\alpha\right)\right)  /\left(1-\alpha\right) - 2+\delta^{\prime}(1) \right)^{2}$. This result will be useful for values of $\alpha$ close to 1.

We now want to illustrate Theorem~\ref{th:ErrorEstimU} with the particular case of a Gumbel copula, whose expression is:
$$C\left(u, v\right)=\exp \left[-\left\{\left(-\ln \left(u\right)\right)^{\theta}+\left(-\ln \left(v\right)\right)^{\theta}\right\}^{\frac{1}{\theta}}\right].$$
The MSE of the nonparametric estimator of the upper TDC is then directly related to the parameter $\theta$, as exposed in the following proposition.

\begin{pro}\label{pro:Gumbel}
In the case of the Gumbel copula of parameter $\theta>1$, the asymptotic variance and squared bias of the nonparametric estimator $\widehat \lambda_{U,n}\left(i(n)/n\right)$ of the upper TDC, defined in equation~\eqref{eq:NonParmLambdaU}, with the assumptions of Theorem~\ref{th:ErrorEstimU}, are:
$$\left\{\begin{array}{ccl}
\text{variance} & = &\frac{1}{n(1-\alpha)^{2}} \left(\delta(\alpha)[1-\delta(\alpha)] + \delta(\alpha)^2\left[\frac{1}{\alpha}-1\right] 2^{\frac{1}{\theta}}\left[2^{\frac{1}{\theta}-1}-2\right] + \delta(\alpha)^2 2^{\frac{2}{\theta}-1}\left[\frac{\delta(\alpha)}{\alpha^2}-1\right]\right)   \\
\text{bias}^2 & = &\left(\frac{1-2 \alpha+\delta\left(\alpha\right)}{1-\alpha} -2+2^{1/\theta} \right)^{2}  ,
\end{array}\right.$$
for $\delta(\alpha)=\exp \left[-\left\{ 2\left(-\ln \left(\alpha\right)\right)^{\theta}\right\}^{\frac{1}{\theta}}\right]$.
\end{pro}

The proof is postponed to Appendix~\ref{sec:proofGumbel}.

\subsection{Average of estimators}

We are now interested in the average estimator:
\begin{equation}\label{eq:AverageNonParmLambdaL}
\widehat{\Lambda}_{L,n}\left(\frac{i_1}{n},...,\frac{i_m}{n}\right)=\frac{1}{m}\sum_{k=1}^{m}{\widehat \lambda_{L,n}\left(\frac{i_k}{n}\right)},
\end{equation}
which is the average of $m$ nonparametric TDC estimators of respective thresholds $i_1,...,i_m\in\llbracket 1,n\rrbracket$. Such an average nonparametric estimator appears for example in the plateau-finding algorithm, which we will describe in Section~\ref{sec:plateau}. It is intended to reduce the MSE of the previously introduced estimators. However, many combinations of $m$ isolated estimators are possible, so that the minimization of the MSE is computationally expensive. For this reason, we put forward a simpler version of this method in the simulation study detailed in Section~\ref{sec:simul}, in which $i_1,...,i_m$ are consecutive numbers. Theorem~\ref{th:ErrorEstimAverage} provides a formula for the MSE of the average estimator.

\begin{thm}\label{th:ErrorEstimAverage}
Let the bivariate copula $C$ have continuous partial derivatives and $i_k(n)$ be equal to $\alpha_k n$, where $\alpha_k\in(0,1)$, for $k\in\llbracket 1,m\rrbracket$, and $m\geq 1$. Then, the MSE of the average nonparametric estimator of the lower TDC, defined in equation~\eqref{eq:AverageNonParmLambdaL}, behaves asymptotically in the following manner:
$$\E\left[\left(\widehat \Lambda_{L,n}\left(\frac{i_1(n)}{n},...,\frac{i_m(n)}{n}\right)-\lambda_L\right)^2\right]  = V^{\Lambda}_{L,n}(\alpha) + \left(\frac{1}{m}\sum_{k=1}^{m}{\frac{1}{\alpha_k}\delta\left(\alpha_k\right)}-\delta'(0)\right)^2,$$
where 
$$\underset{n\rightarrow\infty}{\lim} nV^{\Lambda}_{L,n}(\alpha)= \frac{1}{m^2}\sum_{k,l=1}^{m}{\frac{1}{\alpha_k\alpha_l}\mathcal K\left(\alpha_k,\alpha_l\right)}$$
and
$$\begin{array}{ccl}
\mathcal K(u,v) & = & \delta(u\wedge v)-\delta(u)\delta(v) + (h_1(u)h_1(v)+h_2(u)h_2(v))((u\wedge v)-uv) \\
 & & - h_1(v)(C(u\wedge v,u)-v\delta(u)) - h_2(v)(C(u,u\wedge v)-v\delta(u)) \\
 & & - h_1(u)(C(u\wedge v,v)-u\delta(v)) - h_2(u)(C(v,u\wedge v)-u\delta(v)) \\
 & & + h_1(u)h_2(v)(C(u,v)-uv)+ h_1(v)h_2(u)(C(v,u)-uv)),
\end{array}$$
and where $a\wedge b$ is the minimum between $a$ and $b$.
\end{thm}

The proof is postponed to Appendix~\ref{sec:proofErrorEstimAverage}. The extension of this theorem to an average upper TDC estimator is straightforward and is thus omitted.

According to Theorem~\ref{th:ErrorEstimAverage}, the variance of the average nonparametric estimator of the lower TDC relies on a function $\mathcal K$. If $u=v$, $\mathcal K(u,v)$ is simply equal to $\sigma^2(u)$, where $\sigma^2(u)$ is provided by equation~\ref{eq:sigmaL}. The case of $C$ symmetric also simplifies the expression of $\mathcal K$ :
$$\begin{array}{ccl}
\mathcal K(u,v) & = & \delta(u\wedge v)-\delta(u)\delta(v) + 2h(u)h(v)((u\wedge v)-uv) \\
 & & - 2h(v)(C(u\wedge v,u)-v\delta(u)) - 2h(u)(C(u\wedge v,v)-u\delta(v)) \\
 & & + 2h(u)h(v)(C(u,v)-uv).
\end{array}$$
The function $\mathcal K$ provides some insights into the dependence between two standard nonparametric estimators. More precisely, if we consider the two estimators of different thresholds, $\widehat\lambda_{L,n}(i/n)$ and $\widehat\lambda_{L,n}(j/n)$, we can either write their asymptotic covariance as $2\mathcal K(i/n,j/n)n/ij$ or as $2\rho_{i,j}\sigma(i/n)\sigma(j/n)n/ij$, where $\rho_{i,j}$ is the correlation between $\widehat\lambda_{L,n}(i/n)$ and $\widehat\lambda_{L,n}(j/n)$. As a consequence, the asymptotic correlation $\rho_{i,j}$ is equal to $\mathcal K\left(i/n,j/n\right)/\sigma\left(i/n\right)\sigma\left(j/n\right)$.

\begin{figure}[htbp]
  \centering
  \includegraphics[width=0.7\linewidth]{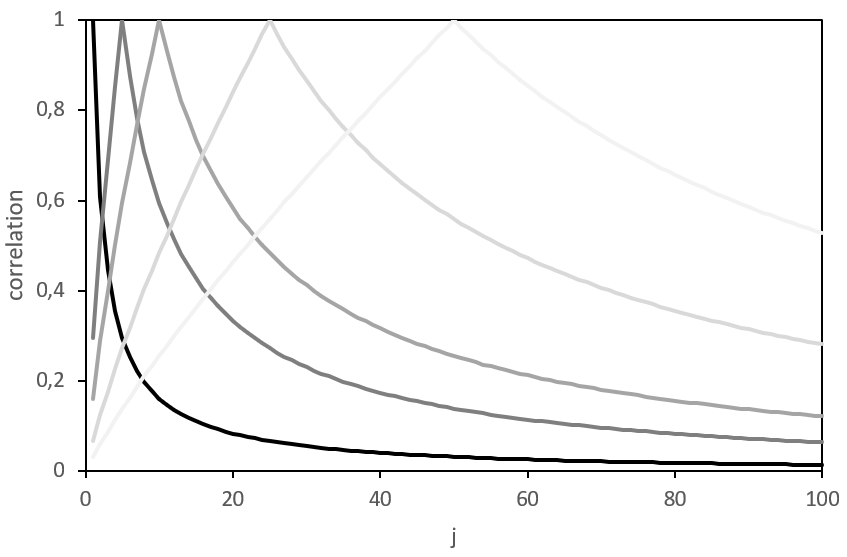}
  \begin{minipage}{0.7\textwidth}\caption{Correlation $\rho_{i,j}$ between $\widehat\lambda_{L,n}(i/n)$ and $\widehat\lambda_{L,n}(j/n)$ for $i$ equal to (from the darkest to the lightest) 1, 5, 10, 25, and 50, and various values of $j$. We consider a Clayton copula and $n=1000$.}\label{fig:correl}
  \end{minipage}
\end{figure}

We see in Figure~\ref{fig:correl} this correlation $\rho_{i,j}$ in the case of the Clayton copula. The smaller $i$, the stronger the correlation decay with respect to $j$. In other words, the impact on an isolated TDC estimator, when one changes the threshold by a fixed amount, is relatively greater when the initial threshold is extreme. This simply illustrates the lack of statistical robustness of estimators relying on few (extreme) observations: a slight expansion of the data taken into account, with respect to the initial number of observations involved in the estimator, may have significant consequences.

\section{Selection of the threshold}

The definition of the nonparametric estimators of the TDC, as in equations~\eqref{eq:NonParmLambdaL}, \eqref{eq:NonParmLambdaU}, and~\eqref{eq:AverageNonParmLambdaL}, relies on a free parameter $i$. A proper estimation of the TDC thus requires an appropriate selection of this free parameter, which we subsequently call the \textit{threshold} since the estimators focus on extreme observations whose rank is beyond the threshold $i$. We first recall a classical selection rule, known as the plateau-finding algorithm~\citep{FJS}. We then propose alternative methods based on the minimization of the asymptotic MSE as expressed in Theorems~\ref{th:ErrorEstim} and \ref{th:ErrorEstimU}. We finally suggest an extension for average estimators.

\subsection{Plateau-finding algorithm}\label{sec:plateau}

Justified by the homogeneity property of the tail copula~\citep{SS2006}, the plateau-finding algorithm is a heuristic algorithm that selects the threshold in a characteristic plateau appearing for estimators $\widehat{\lambda}_{L,n}(i/n)$ or $\widehat{\lambda}_{U,n}(i/n)$ of successive $i$ \citep{FJS}. The algorithm is shown below. Since it works both for the lower and upper tails, we have removed the subscripts $L$ and $U$. In this paragraph, each $\widehat{\lambda}_n(i/n)$ thus refers to equation~\eqref{eq:NonParmLambdaL} or~\eqref{eq:NonParmLambdaU}.
\begin{enumerate}
    \item The series $\{ \widehat{\lambda}_n(i/n)\}_{i\in\llbracket 1,n \rrbracket}$ is smoothed using a box kernel with bandwidth $b \in \mathbb{N}$, which consists in applying a moving average on $2b+1$ consecutive elements. We note $\{ \bar{\lambda}_n(i/n)\}_{i\in\llbracket 1,n-2b \rrbracket}$ the new smoothed series, where $b$ is chosen such that 1\% of the data falls into the box, that is $b=\lfloor n/200\rfloor$.
    \item We want to select a vector $p_{k}=\left(\bar{\lambda}_n(k/n), ..., \bar{\lambda}_n((k+m-1)/n)\right)$ of $m=\lfloor\sqrt{n-2 b}\rfloor$ consecutive estimates, where $k\in\llbracket 1, n-2 b-m+1\rrbracket$. More precisely, the algorithm selects the index $k^{\star}$ of the first\footnote{ Starting from $k=1$ for the lower TDC and from $k=n-2 b-m+1$ for the upper TDC.} vector $p_k$ which satisfies the following plateau condition:
    $$\sum_{i=1}^{m-1}\left|\bar{\lambda}_n((k+i)/n)-\bar{\lambda}_n(k/n)\right| \leq 2 \sigma,$$
    where $\sigma$ is the standard deviation of the smoothed series $\{ \bar{\lambda}_n(i/n)\}_{i\in\llbracket 1,n-2b \rrbracket}$.
    \item Then, the TDC estimator is defined as the average of the estimators $\bar{\lambda}_n(.)$ in the plateau $p_{k^{\star}}$:
    $$
    \check{\lambda}_n=\frac{1}{m} \sum_{i=0}^{m-1} \bar{\lambda}_n((k^{\star}+i)/n).
    $$
\end{enumerate}
If there is no vector fulfilling the plateau condition, the TDC estimate is set to zero.

\subsection{Minimization of the MSE}\label{sec:minMSE}

As an alternative to the plateau-finding algorithm, we propose selecting the threshold minimizing the asymptotic MSE as expressed in Theorems~\ref{th:ErrorEstim} or~\ref{th:ErrorEstimU}, to balance the bias and the variance of the nonparametric TDC estimator. However, minimizing this MSE leads to two issues. The first is that the formulas of the MSE of the nonparametric estimators depend on the true and unobserved copula of the dataset. We can then consider a plug-in approach, in which the unobserved copula is replaced in the MSE formula by an empirical estimate. Nonetheless, this leads to the second issue: in addition to the copula itself, the MSE formula includes derivatives of the copula, namely $\delta'$ and $h$. Regarding this ill-posed inverse problem of estimating derivatives, using a simple empirical estimation of the copula and finite differences, which is an appropriate solution for very large datasets~\citep{GR2004,RS}, may not be enough for smaller ones, so that regularization is then required~\citep{FS}. We thus propose a parametric specification for the unobserved copula, at least for the plug-in in the MSE formulas of the nonparametric estimators of the TDC. 

In this semiparametric approach, we can, for example, assume a Clayton copula when dealing with the lower tail and a Gumbel copula for the upper tail, transforming the MSE formula as in Propositions~\ref{pro:Clayton} and~\ref{pro:Gumbel}. The method we propose is however more general and one may choose parametric copulas other than these traditional examples of tail-dependent copulas. Whether the copula is a Clayton or a Gumbel, it depends on a parameter $\theta$ to be estimated. One could estimate $\theta$ using all observations. However, this approach may be strongly biased. We indeed want this specific parametric copula to depict only the tail of the true copula. We thus propose below two competing methods, in which we focus on extreme observations to estimate $\theta$.

We note that this idea of a plug-in to select the most appropriate free parameter of a nonparametric estimator is a very common practice in nonparametric statistics. It is for example widespread in the literature about kernel density estimation~\citep{JMS}.

\subsubsection{Simple plug-in approach}\label{sec:simpleplug}

We note $\widetilde{MSE}(i,C)$ the MSE provided in Theorem~\ref{th:ErrorEstim} or in the Theorem~\ref{th:ErrorEstimU}. This theoretical asymptotic MSE depends both on the order $i$ used in the nonparametric estimator of the TDC and on the true and unobserved copula $C$, with respect to which the theoretical MSE is calculated. Since the true copula is unknown, we must estimate it in order to estimate the MSE. As explained above, we consider a model in which the tail of the copula, and only its tail, is close to the tail of a specific parametric copula. For example, one can estimate a Clayton copula by considering only extreme observations, so that the estimate is not influenced by the rest of the dependence structure, which may not be consistent at all with a Clayton copula. This idea is useful for estimating tail copulas and the most widespread method in this perspective is the censored likelihood approach~\citep{STC,HW,CCH}. Here, we propose an estimation method combining the nonparametric estimator and the censored likelihood, both depending on a common threshold which separates the extremes from other observations. 

For this purpose, one has to define clearly what are extreme observations. In dimension higher than 1, sorting vectors and thus defining extreme vectors and quantile vectors is a question for which one can conceive several different solutions, such as spatial quantiles~\citep{AT}, geometric quantiles~\citep{Chaudhuri}, or quantiles based on the inversion of an appropriate mapping~\citep{Koltchinskii,GGH}. Following this last idea, we define here an extreme observation as one belonging to the empirical orthant quantile of probability lower than $i/n$. In practice, we first determine the probability $\widehat F_n$ associated with each observation $(X_j,Y_j)$, that is the empirical probability to have an observation in the lower left orthant of $(X_j,Y_j)$. Then, for a given threshold $i$, the set of corresponding observations is defined by 
$$\Omega^L_{i/n}=\left\{(X_j,Y_j)\in\mathbb R^2,j\in\llbracket 1,n\rrbracket \left| \widehat{F}_n\left(X_{j},Y_{j}\right)\leq \widehat C_n\left(\frac{i}{n},\frac{i}{n}\right) \right.\right\}$$
for the lower tail and
$$\Omega^U_{i/n}=\left\{(X_j,Y_j)\in\mathbb R^2,j\in\llbracket 1,n\rrbracket \left| \widehat{F}_n\left(X_{j},Y_{j}\right)\geq \widehat C_n\left(\frac{i}{n},\frac{i}{n}\right) \right.\right\}$$
for the upper tail.
%, with $(X_{j},Y_j)$ an observation and $\widehat{F}_X$ and $\widehat{F}_Y$ the corresponding empirical marginals. Each of these sets contains between $0$ and $i$ observed vectors, depending on the dependence between the two components. In the simulations and applications of this paper, we will work with positive dependencies between the components, so that we do not encounter empty $\Omega^L_{i/n}$ and $\Omega^U_{i/n}$ in practice.

We note that the probability for a pair of observations to be in $\Omega^L_{i/n}$ is $K(C(i/n,i/n))$, where $K$ is the Kendall function associated with the probability distribution $F$: $K(p)=\proba[F(X,Y)\leq p]$. We justify this Kendall quantile approach by the fact that a vector $(X_j,Y_j)$ dominates all the observations whose probability is lower than $\widehat F_n(X_j,Y_j)$ and not only those in the lower left quadrant of $(X_j,Y_j)$~\cite{GGH}. The Kendall function will be overriding for calculating censored likelihoods. It is worth noting that the Kendall function is unique for a given copula and that its expression is straightforward in the case of Archimedean copulas, namely it is $K_{\theta}(p)=p-p\ln(p)/\theta$ for the Gumbel copula and $K_{\theta}(p)=p+p^2(1-p^{\theta})/\theta$ for the Clayton copula~\citep{BGGR,GR2001,GGH}.

Given a threshold $i$, one estimates a parametric copula close to the true copula of the extreme vectors by a censored maximum likelihood method, restricted to the observations either in $\Omega^L_{i/n}$ or in $\Omega^U_{i/n}$. We note $c_{\theta}$ and $C_{\theta}$ the parametric copula density and cumulative distribution function, which are not specified more precisely here\footnote{ As previously explained, this copula may for example be a Clayton copula for the lower tail or a Gumbel copula for the upper tail. We can even work with a set of various parametric copulas and finally select the pair of copula specification and parameter with the highest likelihood or the highest AIC/BIC. This extension, though promising, is not developed further in this paper. In particular, it implies an explicit formula for the MSE of each specification of parametric copula, as we do for Clayton and Gumbel copulas.} and which are parameterized by $\theta$. We also note $K_{\theta}$ the parametric Kendall function corresponding to the copula $C_{\theta}$. In this censored approach, the considered likelihood is $c_{\theta}(\widehat{F}_{X,n}(X_j),\widehat{F}_{Y,n}(Y_j))$ for any vector in the set of extreme observations $\Omega^L_{i/n}$, whereas we replace this likelihood by the probability measure of the set $\mathbb R^2\setminus\Omega^L_{i/n}$ for any non-extreme observation. In order to take into account the parameter $\theta$ in this last probability, we consider a pseudo probability, where only the Kendall function depends on $\theta$. Therefore, with this assumption, the probability measure of $\mathbb R^2\setminus\Omega^L_{i/n}$ is $1-K_{\theta}(\widehat F_n(X_j,Y_j))$, where $(X_j,Y_j)$ is on the boundary of the set $\Omega^L_{i/n}$. The estimator $\widehat\theta_{L,i/n}$ of $\theta$ is thus, for the lower tail:
$$\widehat\theta_{L,i/n}=\underset{\theta}{\text{argmax}} \sum_{j=1}^{n}{\left\{\ln\left(c_{\theta}(\widehat{F}_{X,n}(X_j),\widehat{F}_{Y,n}(Y_j))\right) \mathcal J^L_{j,i} + \ln \left(1-K_{\theta}\left(\widehat C_n\left(\frac{i}{n},\frac{i}{n}\right)\right)\right)\left(1-\mathcal J^L_{j,i}\right)\right\}},$$
where $\mathcal J^L_{j,i}=\indic_{(X_j,Y_j)\in\Omega^L_{i/n}}$. We have a similar formula for the upper tail:
$$\widehat\theta_{U,i/n}=\underset{\theta}{\text{argmax}} \sum_{j=1}^{n}{\left\{\ln\left(c_{\theta}(\widehat{F}_{X,n}(X_j),\widehat{F}_{Y,n}(Y_j))\right) \mathcal J^U_{j,i} + \ln \left(K_{\theta}\left(\widehat C_n\left(\frac{i}{n},\frac{i}{n}\right)\right)\right)\left(1-\mathcal J^U_{j,i}\right)\right\}},$$
where $\mathcal J^U_{j,i}=\indic_{(X_j,Y_j)\in\Omega^U_{i/n}}$.

In both formulas, the likelihood is in fact a pseudo likelihood insofar as it uses the empirical marginal distributions instead of a parametric specification with parameters to be estimated. In the simulation study, we even work directly with pseudo observations, insofar as we simulate random variables with a uniform marginal distribution. This approach is a common practice for estimating copulas and leads to an estimator of copula parameters which is asymptotically normal and consistent \citep{genest1995,shih1995}. Apart from this consideration, the estimates $\widehat\theta_{L,i/n}$ and $\widehat\theta_{U,i/n}$ depend on the choice of the threshold $i$ and we can naturally define a mapping $\psi$ such that $\widehat\theta_{L,i/n}=\psi(i/n)$, or $\widehat\theta_{U,i/n}=\psi(i/n)$ if we are instead interested in the upper tail.

Using a plug-in approach, the MSE is now estimated by $\widetilde{MSE}(i,C_{\psi(i/n)})$, which can be expressed thanks to Proposition~\ref{pro:Clayton} (respectively Proposition~\ref{pro:Gumbel}) if we use the Clayton (resp. Gumbel) specification for lower (resp. upper) tails. The optimal $i$ in this plug-in approach is then:
$$i^{\star}_{PI}=\underset{i\in\llbracket 1,n\rrbracket}{\text{argmin}}\  \widetilde{MSE}(i,C_{\psi(i/n)}).$$
In other words, given a threshold $i$, we calculate the corresponding estimated MSE between the nonparametric TDC using this threshold and a TDC for a parametric copula, which we estimate on observations beyond the same threshold $i$. Finally, we select the threshold $i^{\star}_{PI}$ minimizing this MSE.

\begin{figure}[htbp]
  \centering
  \includegraphics[width=0.7\linewidth]{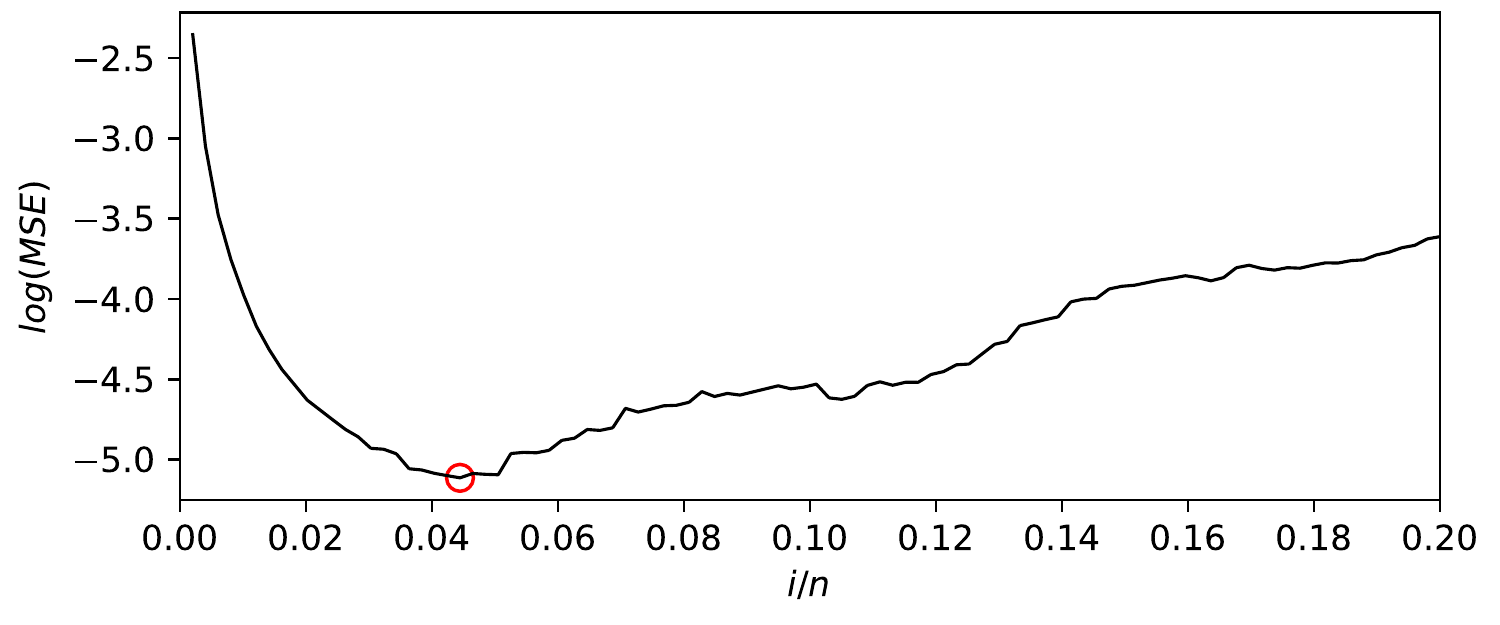} 
  \begin{minipage}{0.7\textwidth}\caption{Function of $\log\left(\widetilde{MSE}(i,C_{\psi(i/n)})\right)$ with respect to the threshold $i/n$, for $n=1000$ pairs simulated with a rotated Gumbel copula of parameter $1.5$.}\label{fig:plugin}
  \end{minipage}
\end{figure}

The TDC estimator is then the standard nonparametric TDC estimator for the selected threshold:
$$\widehat{\lambda}_{L,MSE,n}=\widehat{\lambda}_{L,n}\left(\frac{i_{PI}^{\star}}{n}\right),$$
where $L$ has to be replaced by $U$ for the upper case. Figure~\ref{fig:plugin} illustrates the selection of this threshold in the case of a lower tail, with data simulated by a rotated Gumbel copula.

\subsubsection{Two-step plug-in approach}\label{sec:twostep}

In the simple plug-in approach described above, we have selected the rank $i^{\star}_{PI}$ minimizing the estimated MSE, either for the lower or for the upper TDC. However, if we consider that $C_{\psi(i^{\star}_{PI}/n)}$ is the best estimation of the true copula $C$, at least in the tail of the distribution, one could then argue that the rank $i^{\star}_{PI}$ is not necessarily the one minimizing the estimated MSE of the TDC estimator: in other words, one may find a rank $i$ such that $\widetilde{MSE}(i,C_{\psi(i^{\star}_{PI}/n)}) < \widetilde{MSE}(i^{\star}_{PI},C_{\psi(i^{\star}_{PI}/n)})$. We are thus eager to find a rank $i$ such that it minimizes the MSE of the TDC estimator with the copula $C_{\psi(i/n)}$: this rank $i$ must verify the following fixed-point equation:
\begin{equation}\label{eq:i2_argmin}
i=\underset{j\in\llbracket 1,n\rrbracket}{\text{argmin}}\  \widetilde{MSE}(j,C_{\psi(i/n)}).
\end{equation}
This objective leads to a two-step plug-in approach described below.

%\textcolor{red}{Autre motivation : l'estimation de $\theta$ est très incertaine, surtout quand on prend un $i$ extrême, donc estimation de MSE en souffre. En imposant de minimiser la MSE pour chaque theta estimé, on s'impose de ne pas être dans une zone où $i$ est trop extrême et génère une variance importante sur les estimateurs. Concrètement, le two-step plug-in va améliorer le plug-in simple.}

\begin{figure}[htbp]
  \centering
  \includegraphics[width=0.95\linewidth]{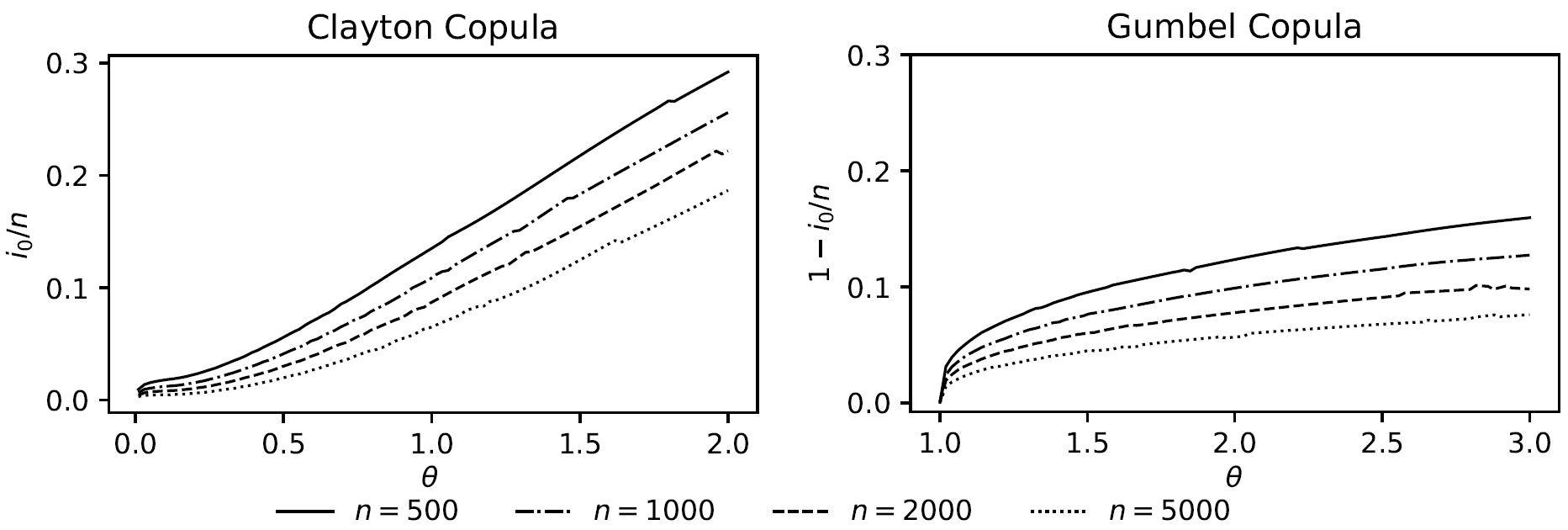} 
  \begin{minipage}{0.7\textwidth}\caption{Optimal threshold $\phi(\theta)$ (respectively $1-\phi(\theta)$) of the lower (resp. upper) TDC estimator, given the copula parameter $\theta$, for various values of $n$, for the Clayton (resp. Gumbel) copula.}\label{fig:mse}
  \end{minipage}
\end{figure}

From Theorem~\ref{th:ErrorEstim}, given a copula $C_{\theta}$, we can determine the threshold $i_0$ minimizing the theoretical asymptotic MSE of the nonparametric TDC estimator. We can thus define a mapping $\phi$ between the parameter $\theta$ and the corresponding optimal threshold in the nonparametric TDC estimator: $i_0/n=\phi(\theta)$. The formal definition of $\phi$ is as follows:
\begin{equation}\label{eq:phi_argmin}
\phi(\theta)=\frac{1}{n}\ \underset{j\in\llbracket 1,n\rrbracket}{\text{argmin}}\  \widetilde{MSE}(j,C_{\theta}).
\end{equation}
If we plot this function $\phi$ in the particular case of a Clayton or a Gumbel copula, we observe a strictly monotonic function, as one can see in Figure~\ref{fig:mse}. In these cases, we can numerically invert $\phi$. In a broader perspective, we can define the generalized inverse function, $\phi^{-1}:u\in[0,1]\mapsto\inf\{\theta\in\mathbb R, \phi(\theta)\geq u\}$, and finally write $\theta=\phi^{-1}(i_0/n)$.

The $\theta$ parameter being unknown, we now replace it by its estimator $\widehat\theta_{L,i/n}$. As stated in equation~\eqref{eq:i2_argmin}, we are looking for a rank $i$ defining the same threshold for the estimation of $\theta$ and for the estimation of the TDC. Therefore, combining equations~\eqref{eq:i2_argmin} and~\eqref{eq:phi_argmin} leads to the following equation for the optimal $i$:
$$\frac{i}{n}=\frac{1}{n}\ \underset{j\in\llbracket 1,n\rrbracket}{\text{argmin}}\  \widetilde{MSE}(j,C_{\psi(i/n)})=\phi\left(\psi\left(\frac{i}{n}\right)\right).$$
In other words, our objective is to have 
\begin{equation}\label{eq:phipsi}
\psi\left(\frac{i}{n}\right)=\phi^{-1}\left(\frac{i}{n}\right).
\end{equation}
We thus define the optimal threshold $i$ as the rank for which the graphs of $\psi$ and $\phi^{-1}$ intersect themselves. Depending on the observations and on the parametric copula, this intersection may not exist or be multiple. Moreover, the discrete nature of the threshold makes highly improbable the existence of a fixed point, that is of a $i$ satisfying equation~\eqref{eq:phipsi}. Therefore, we instead minimize the quadratic deviation between $\psi(i/n)$ and $\phi^{-1}(i/n)$, so that our estimated optimal threshold $i_{2PI}^{\star}$ is such that:
$$i_{2PI}^{\star}=\underset{i\in\llbracket 1,n\rrbracket}{\text{argmin}}\left(\psi\left(\frac{i}{n}\right)-\phi^{-1}\left(\frac{i}{n}\right)\right)^2.$$

In practice, the determination of $i_{2PI}^{\star}$ thus amounts to an optimization algorithm. We can for instance propose to use Nelder-Mead's algorithm \citep{NM}. In this case, we start with two different and arbitrary thresholds, for which we determine the output of the objective function $i\mapsto\left(\psi(i/n)-\phi^{-1}(i/n)\right)^2$. Then, the iteration rule of Nelder-Mead makes these two thresholds evolve and finally converge towards a local minimum. Refinements, such as the mix of several executions of this algorithm, could improve the results and lead to reach a threshold closer to a global minimum.

\begin{figure}[htbp]
  \centering
  \includegraphics[width=0.75\linewidth]{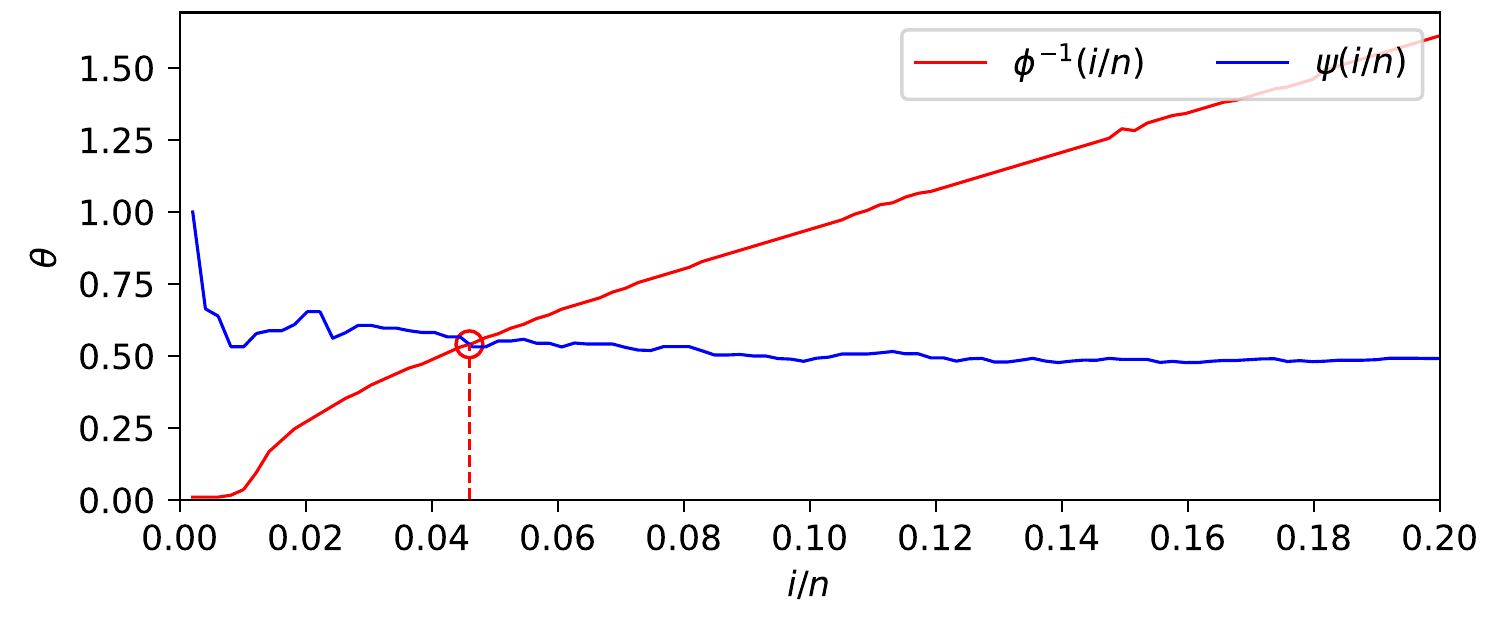} 
  \begin{minipage}{0.7\textwidth}\caption{Threshold selection for the two-step plug-in approach for the lower TDC. The number $n$ of simulated pairs is equal to 1000, and the data-generating copula is a rotated Gumbel copula of parameter $1.5$. The $\theta$ at the ordinate is the parameter of the Clayton copula used by our estimator to describe the lower tail dependence.}\label{fig:twosteps}
  \end{minipage}
\end{figure}

This heuristic approach provides satisfying results in simulations. Figure~\ref{fig:twosteps} illustrates the principle of the selection of the optimal threshold, corresponding to the abscissa of the intersection of the two curves. In this example, the optimal threshold is lower with the two-step plug-in estimator than with the simple plug-in illustrated in Figure~\ref{fig:mse}.

%\textcolor{red}{Si optimum multiple, on prend le plus grand i pour diminuer variance.}

Similarly to the simple plug-in approach, the TDC estimator in this two-step plug-in is the nonparametric TDC estimator for the selected threshold:
$$\widehat{\lambda}_{L,MSE2,n}=\widehat{\lambda}_{L,n}\left(\frac{i_{2PI}^{\star}}{n}\right),$$
where $L$ has to be replaced by $U$ for the upper tail.

\subsection{Minimizing an average MSE}\label{sec:averageest}

The plateau algorithm leads to estimations of the nonparametric TDC with a particularly low variance. We can explain this characteristic of this estimation method by the double regularization of the TDC which is performed in steps 1 and 3 of the algorithm detailed in Section~\ref{sec:plateau}. Inspired by this regularization in the plateau algorithm, we can also propose an additional regularization of the MSE-based plug-in estimators. 

The smoothing procedure we propose is similar to step 3 of the plateau algorithm. We do not mimic step 1, which consists of first smoothing the estimated nonparametric TDC, because this step would modify the distribution of this TDC and make our MSE estimate erroneous.

We note that the smoothing used in the plateau is a simple rule-of-thumb averaging. It could be beneficial to use other smoothing techniques based on the minimization of the error induced by smoothing. Among these techniques, which are omnipresent in nonparametric statistics~\citep{HSMW}, one can cite smoothing with wavelets~\citep{Mallat2,Ranta,GG3,GGarch} or smoothing resulting from a variational problem~\citep{Garcin2017}. We will not use these methods here, to be congruent with the plateau algorithm and to make fair comparisons between the various TDC estimators. Nevertheless, we will see that beyond the arbitrary averaging, one can also do an averaging minimizing the MSE, as a result of Theorem~\ref{th:ErrorEstimAverage}.

We are given a parameter $m$ describing the size of an interval $\mathcal I_m$ of consecutive ranks, where $m$ is the size of the plateau in the algorithm described in Section~\ref{sec:plateau}. We now have to select an appropriate interval $\mathcal I_m$ of consecutive ranks, so that our new regularized TDC estimator will be:
$$\widehat{\lambda}_{L,\mathcal I_m,n}=\frac{1}{m}\sum_{i\in\mathcal I_m}{\widehat{\lambda}_{L,n}\left(\frac{i}{n}\right)},$$
where $L$ has to be replaced by $U$ for the upper tail. 

We can propose several ways of selecting $\mathcal I_m$. For example, from the knowledge of an optimal plug-in rank, $i^{\star}_{PI}$ or $i^{\star}_{2PI}$, we can build $\mathcal I_m$ as an interval having this optimal rank in its median or in one of its bounds, such as $\llbracket i^{\star}_{2PI},i^{\star}_{2PI}+m-1\rrbracket$.

Alternatively, we can select an interval of ranks minimizing the average of the MSE of each rank. More precisely, for each rank $i$, we are able to approximate the corresponding MSE of the nonparametric TDC estimator, following the simple plug-in approach: $\widetilde{MSE}(i,C_{\psi(i/n)})$. We now select the $m$ consecutive ranks leading to the minimal average estimated MSE. We note $k^{\star}_{PI}$ the left bound of this interval of $m$ ranks:
$$k^{\star}_{PI}=\underset{k\in\llbracket 1,n-m+1\rrbracket}{\text{argmin}}\  \frac{1}{m}\sum_{i=1}^{m}{\widetilde{MSE}(k+i-1,C_{\psi((k+i-1)/n)})}.$$
The resulting estimated TDC is the average of the TDC estimates whose rank is in the interval $\mathcal I_m=\llbracket k^{\star}_{PI},k^{\star}_{PI}+m-1\rrbracket$.

Finally, we can also select $\mathcal I_m$ using Theorem~\ref{th:ErrorEstimAverage}. Indeed, given $\mathcal I_m$, this theorem makes it possible to calculate directly the MSE of the average estimator instead of the average MSE of isolated estimators. We can thus extend naturally the direct and two-step plug-in approaches, in which we replace the mappings $\psi$ and $\phi$ respectively by $\Psi$ and $\Phi$, which are defined as follows. Given an interval of ranks $\mathcal I_m$, the mapping $\Psi$ provides an average estimator of the copula parameter, by focusing on each extreme set $\Omega^{L}_{i/n}$ corresponding to each rank $i$ of the interval $\mathcal I_m$:
$$\Psi\left(\mathcal I_m\right)=\frac{1}{m}\sum_{i\in\mathcal I_m}{\psi\left(\frac{i}{n}\right)}.$$
Given a copula parameter $\theta$, $\Phi$ provides the interval $\mathcal I_m=\llbracket i^{\star},i^{\star}+m-1\rrbracket$ minimizing the MSE of an average estimator $\widehat{\Lambda}_{L,n}$ of the TDC:
$$i^{\star}=\underset{i\in\llbracket 1,n-m+1\rrbracket}{\text{argmin}}\   \E\left[\left(\widehat \Lambda_{L,n}\left(\frac{i}{n},...,\frac{i+m-1}{n}\right)-\lambda_L\right)^2\right].$$
This MSE is expressed in Theorem~\ref{th:ErrorEstimAverage}, which can be applied for example with a Clayton copula.

\section{A simulation study}\label{sec:simul}

We compare the estimators introduced above with other common TDC estimators. More precisely, the four estimators based on the minimization of an MSE include a plug-in approach in which we estimate $\theta$ on the whole dataset (as evoked in the preamble of Section~\ref{sec:minMSE}), along with other approaches in which we estimate $\theta$ using only tail data, which include the simple plug-in (Section~\ref{sec:simpleplug}), the two-step plug-in (Section~\ref{sec:twostep}), and a simple plug-in average estimator (Section~\ref{sec:averageest}). We focus our analysis on the upper TDC, where the tail dependence function is given by the Gumbel copula. %Since the upper TDC of copula is equal to the lower TDC of survival copula the results are similar as with the estimation of the lower TDC.

We compute the empirical bias and standard deviation $\sigma(\widehat{\lambda}_{U,n})$ of the estimator for $N=100$ random sample replications of three different sample sizes $n\in\{500, 2000, 5000\}$. We also compute the root-mean-square error (RMSE) of the estimator to analyse the trade-off between bias and variance for all estimation methods: 
$$ \operatorname{RMSE}\left(\widehat{\lambda}_{U,n}\right)=\sqrt{\frac{1}{N} \sum_{j=1}^{N}\left(\widehat{\lambda}_{U,n}^j-\lambda_U\right)^{2}}.$$

We compare the estimators using random data of different samples generated by four different bivariate distributions. We first use a Gumbel copula with parameter values $\theta \in \{ 1.1, 1.5, 1.75, 2 \}$, corresponding to $\lambda_U \in \{ 0.12, 0.41, 0.51, 0.59\}$. The second generated distribution is a bivariate standard t-distribution with $\nu \in \{1, 2, 3\}$ degrees of freedom for correlation values $\rho \in \{0, 0.25\}$, which correspond to six possible TDCs, $\lambda_U \in \{ 0.29, 0.18, 0.12, 0.39, 0.27, 0.20\}$. Third, we generate a distribution with a survival Clayton copula with parameter values $\theta \in  \{0.1, 0.5, 1, 1.5\} $, corresponding to $\lambda_U \in \{ 0, 0.25, 0.50, 0.63\}$. Finally, we focus on a case with tail independence ($\lambda_U = 0 $) corresponding to a Gaussian distribution with correlations $\rho \in \{0, 0.25, 0.5, 0.75 \}$.

For convenience, Table~\ref{tab:estmet} reports the identification number for each of the eight estimators implemented in this study. The arbitrary threshold selection (1) and (2) serve as baseline indicators. We implement also the maximum likelihood estimator (3), The plateau-finding algorithm (4), whereas the other estimators (5) and (6) are the proposed methods based on minimization of the theoretical MSE.

\begin{table}[htbp]
\resizebox{\textwidth}{!}{%
\begin{tabular}{cl}
\hline
Method & Description                                                                               \\ \hline
(1)    & Arbitrary choice of the threshold = 1\%                                                   \\
(2)    & Arbitrary choice of the threshold = 2\%                                                   \\
(3)    & Maximum likelihood estimation with an arbitrary copula function (Gumbel)                  \\
(4)    & Plateau-finding algorithm                                             \\

(5)    & Minimization of the MSE: Simple plug-in estimator                                          \\
(6)    & Minimization of the MSE: Two-step plug-in estimator \\ \hline
\end{tabular}}
\caption{Estimation methods.}\label{tab:estmet}
\end{table}

\subsection{Gumbel simulations}\label{sec:GumSimu}

In this case, we assume that the underlying distribution function is known. For estimator (3), we use a maximum likelihood estimation of the Gumbel copula distribution which is the true sample distribution. For the two estimators introduced in this paper (5-6), the function of the upper TDC estimator is also based on the Gumbel distribution.

The results are gathered in Figure~\ref{fig:gumbsimul}, with more details in Table~\ref{tab:gumb}, in the appendix.

For the lowest value of the true TDC ($\lambda_U=0.12$), that is, when the Gumbel copula has a parameter $\theta = 1.1$, all the estimators show almost similar results in terms of RMSE. In this case, even if the methods (1) and (2) exhibit the lowest bias they have the highest variances. However, for the three other datasets, that is for $\theta \in \{1.5, 1.75, 2\}$, when the true TDC is higher, the methods (1) and (2) are the worst performing estimators in terms of bias, variance and RMSE.

The plateau-finding algorithm (4) and the two proposed methods (5) and (6) have good performance, with a slightly lower bias and variance overall for methods (5) and (6).

The method relying on the maximum likelihood estimation (3) is the best performing estimator overall: not surprisingly, the estimator based on the estimation of the true copula performs better than the others.

\begin{figure}[htbp]%\label{fig:1}
  \centering
  \includegraphics[width=0.95\linewidth]{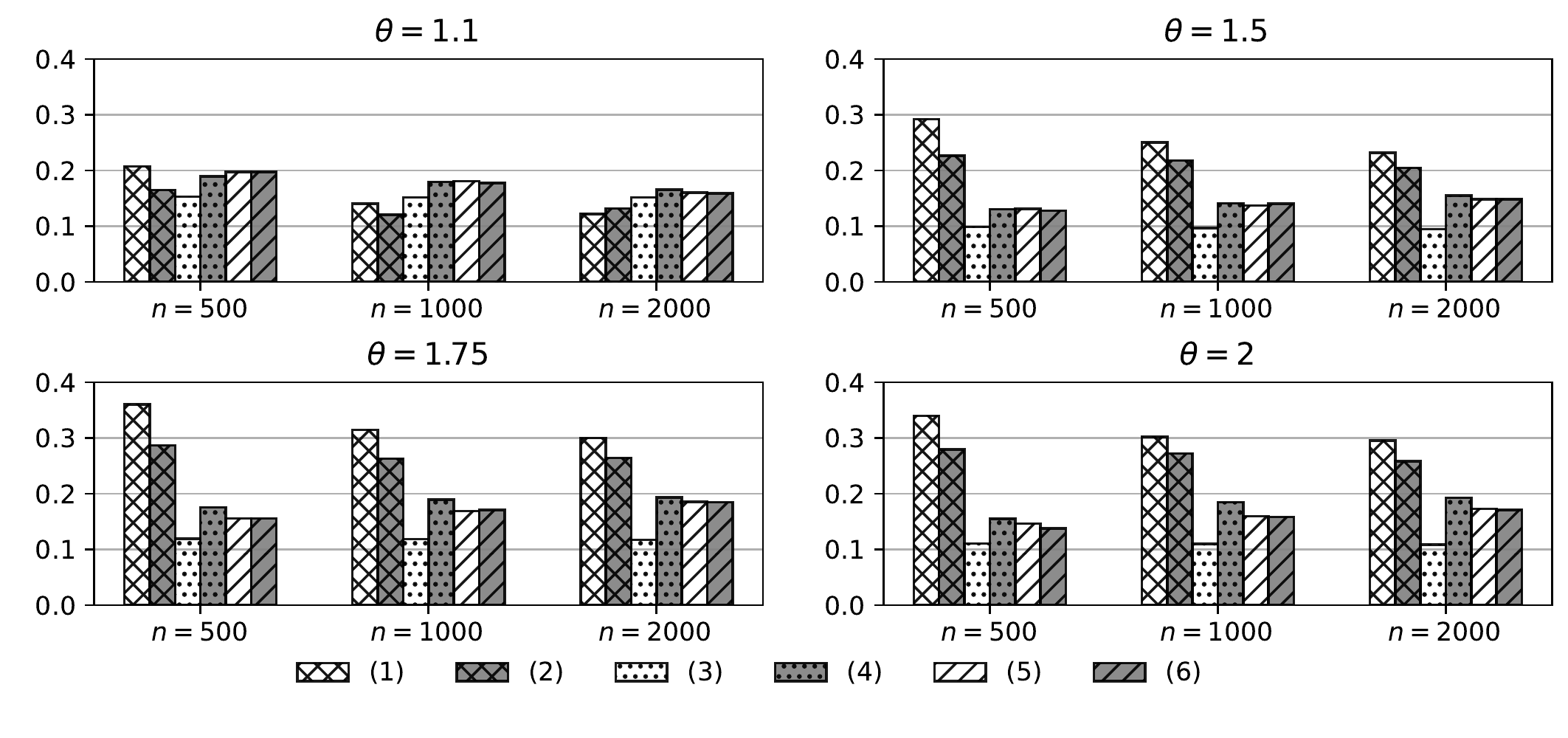} 
  \caption{ RMSE for the upper tail dependence with 100 Gumbel simulations. \label{fig:gumbsimul}}
\end{figure}

\subsection{Student simulations}\label{sec:StuSimu}

Following \cite{SS2006}, we test our estimator on random generations from a bivariate standard t-distribution with $\nu=1, 2, 3$ degrees of freedom. We consider the case with no correlation, $\rho = 0$, and the case where there is a small correlation coefficient, $\rho = 0.25$.

The results are gathered in Figure~\ref{fig:studentsimul}, with more details in Tables~\ref{tab:student} and~\ref{tab:studentRho}, in the appendix. Excepted for estimator (3), the larger the sample size is, the lower the RMSE. Consistent with the findings of \cite{SS2006}, the plateau algorithm (4) performs well regardless of the parameters of the generating model. However we observe quite similar results for estimators (5) and (6) (minimization of the MSE).

For the method based on the maximum likelihood (3) the performance is strongly dependent on the parameterizations considered. It is the best performing estimator for $\rho=0.25$ and $\nu \in \{2, 3\}$, but it is the worst performing estimator when $\rho=0$ and $\nu \in \{1, 2\}$. The performance for the first two estimators (1) and (2) are also quite dependent on the dataset considered and are not performing well overall, compared to the other estimators.

\begin{figure}[htbp]%\label{fig:1}
  \centering
  \includegraphics[width=0.95\linewidth]{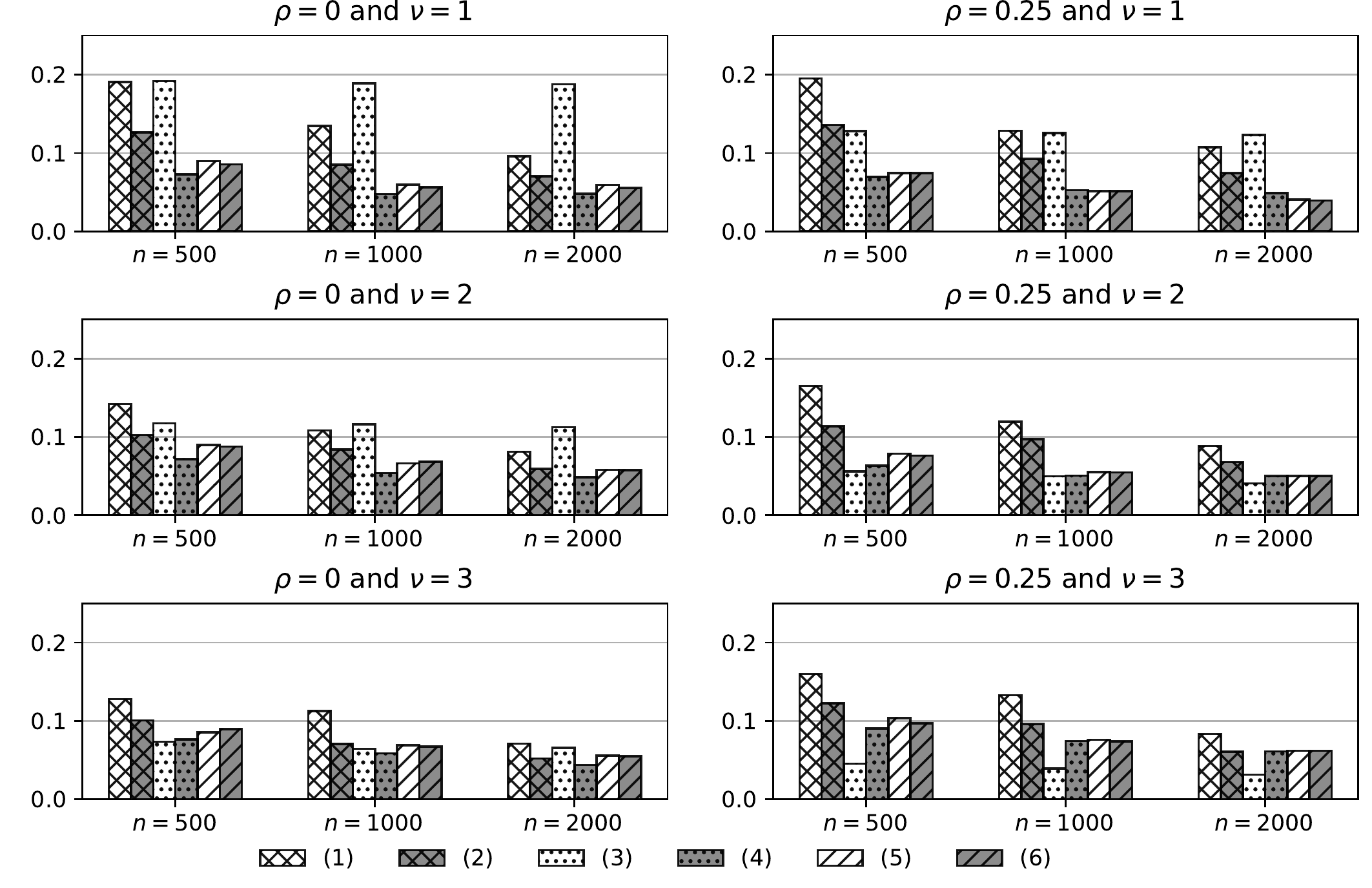} 
    \caption{RMSE for the upper tail dependence with 100 Student simulations. \label{fig:studentsimul}}
\end{figure}

\subsection{Rotated Clayton simulations}

For these simulations, the results are gathered in Figure~\ref{fig:claytonsimul}, with more details in Table~\ref{tab:clayton}, in the appendix.

For the lowest theta parameter ($\theta = 1.1$) when the true TDC is 0, all the estimators exhibit similar results in terms of RMSE, which is also observed in the Gumbel simulations case. The estimators (4), (5), and (6) also behave similarly and perform relatively well across all the parameterizations considered.

\begin{figure}[htbp]%\label{fig:1}
  \centering
  \includegraphics[width=0.95\linewidth]{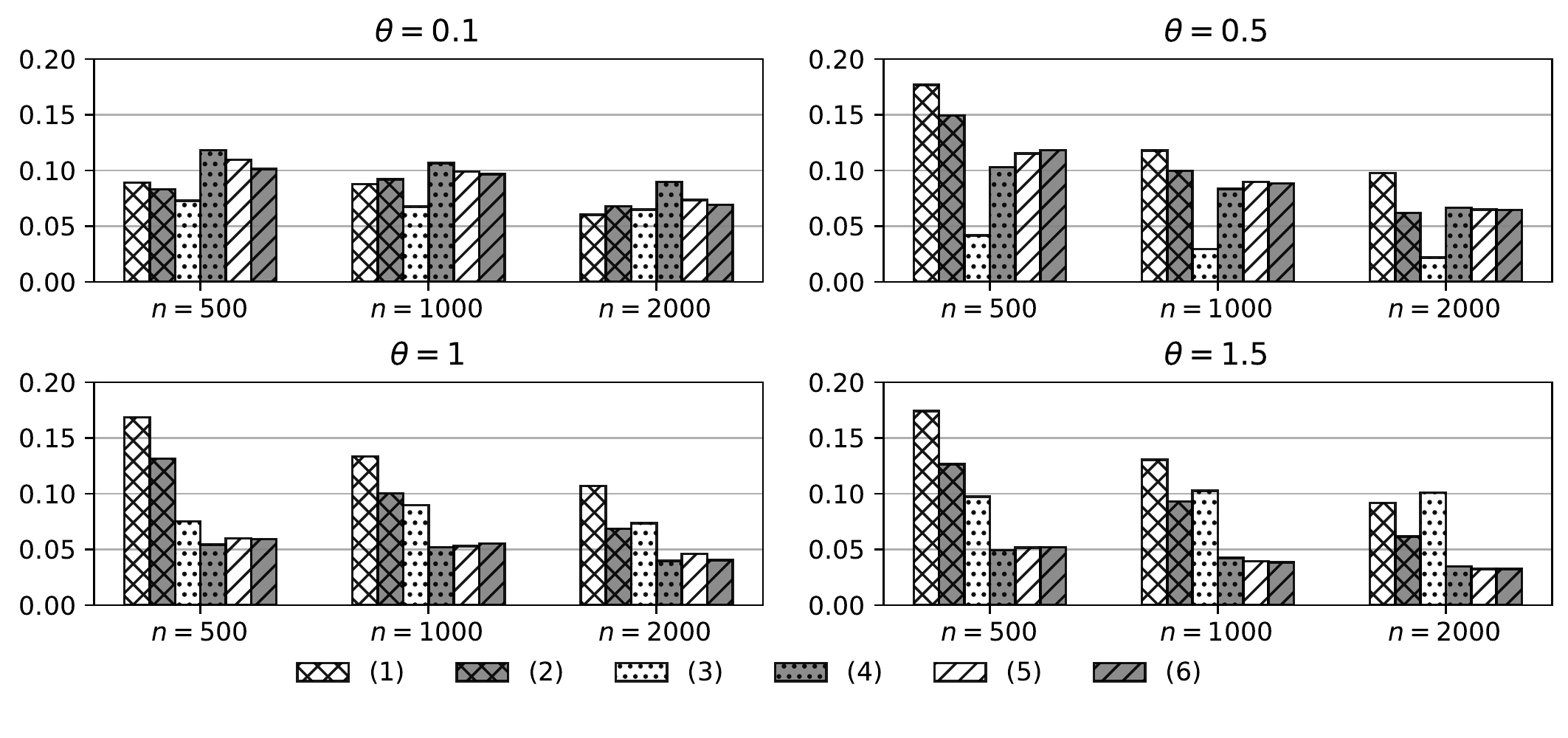} 
  \caption{RMSE for the upper tail dependence with 100 rotated Clayton simulations. \label{fig:claytonsimul}}
\end{figure}

\subsection{Gaussian simulations}

We now evaluate the estimators on the Gaussian copula, that is in a framework with no tail dependence. We see that an increase in the correlation coefficient strongly biases all the estimators. Our results show that the nonparametric estimator for different threshold values captures tail dependence even when the true distribution does not exhibit tail dependence but only dependence for the bulk of the bivariate distribution. This is entirely consistent with \cite{FJS}.

The results are gathered in Figure~\ref{fig:gausssimul}, with more details in Table~\ref{tab:gauss}, in the appendix.

\begin{figure}[htbp]%\label{fig:1}
  \centering
  \includegraphics[width=0.95\linewidth]{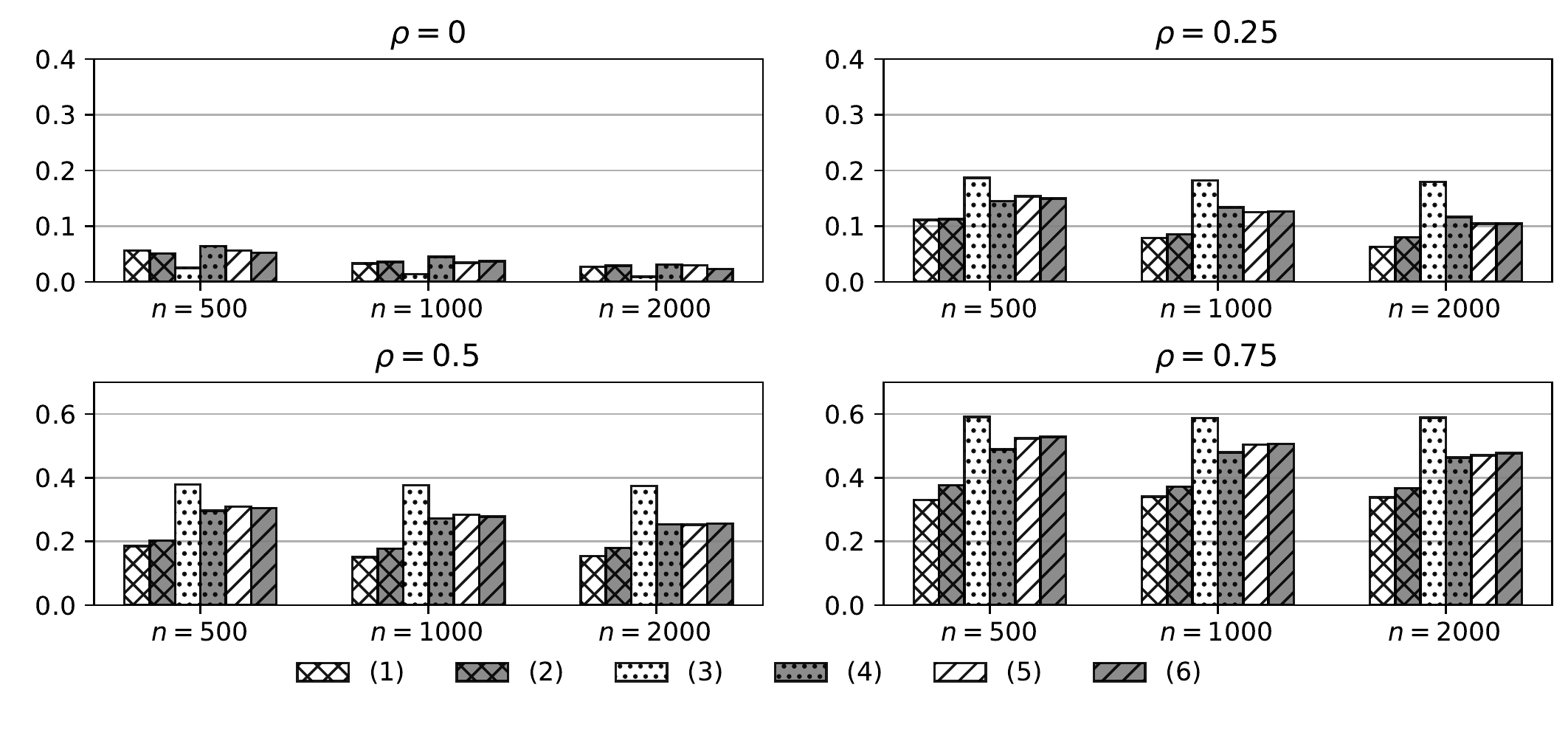} 
  \caption{ RMSE for the upper tail dependence with 100 Gaussian simulations. \label{fig:gausssimul}}
\end{figure}

\subsection{Analysis}

Overall, our two estimators based on a minimization of the theoretical MSE, that is estimators (5) and (6), are, with the plateau-finding algorithm (4), the best performing estimators. They are the least sensitive to a change of parameter in the generating distribution. In contrast, the performance of other estimators strongly depends on the distribution considered and on the sample size. %For several of the examples above, we observe a slightly better performance for our introduced estimators with respect to the plateau-finding method, when $n$ is large. 

Interestingly enough, the maximum-likelihood approach (3) is much less performing when the copula is unknown and differs from Gumbel's one, as discussed in Section~\ref{sec:StuSimu}, than in the oracular Gumbel framework, outlined in Section~\ref{sec:GumSimu}. Conversely, estimators (5) and (6), which are also based on the Gumbel copula, do not seem to be too sensitive to the choice of the parametric copula, presumably because of the censored estimation.

We find very close performances for the plateau-finding algorithm (4) and the MSE-based approaches (5) and (6). However, methods (5) and (6) use the basic nonparametric TDC estimator of equations~\eqref{eq:NonParmLambdaL} and~\eqref{eq:NonParmLambdaU}, whereas the plateau-finding algorithm incorporates additional smoothing stages and thus leads to a more sophisticated nonparametric estimator of the TDC. This simulation study thus highlights the relevance of the estimators based on a minimization of the theoretical MSE, when one has in mind the simple objective of selecting an appropriate threshold in the basic nonparametric estimators of equations~\eqref{eq:NonParmLambdaL} and~\eqref{eq:NonParmLambdaU}, without attempting to refine the expression of this estimator.

However, when the true TDC is close to zero, most estimators, including the plateau-finding algorithm, perform poorly in terms of bias and RMSE, while their variance is relatively low. This result is consistent with the findings of \cite{FJS} and \cite{poulin2007importance}. Therefore, we suggest testing the tail dependence before computing an estimation of the TDC \citep{ledford1996, caperaa1997, hoga2018structural}.

\section{Empirical application}

Copulas are used in many fields, such as hydrology~\citep{tawn1988,GF,poulin2007importance,aghakouchak2010estimation}, astronomy~\citep{SBMB,SIT}, telecommunication networks~\citep{GG1,NHGS}. We focus here on the modelling of financial assets, whose literature also largely uses copulas. In particular, there is a prevalent use of TDCs to describe the dependence of extreme financial returns \citep{malevergne2003testing,poon2004extreme,CG}.

In this short empirical application, we estimate the lower and upper TDCs using the six same estimators as used in the simulation study. We consider the MSCI developed markets indices, which represent the performance of the overall financial securities in countries with a developed market. The estimation period is between 01/01/2000 and 01/01/2021. It consists of 5,295 daily observations for 18 countries. We estimate the pairwise TDCs between the US market and other developed markets. The results are reported in Tables~\ref{tab:lowerUS} and~\ref{tab:upperUS}, along with a standard linear dependence measure, namely the correlation coefficient.

\begin{table}[htbp]
\centering
\begin{tabular}{@{}l|llllll|l@{}}
\hline
Country     & (1)  & (2)  & (3)  & (4)  & (5)  & (6)  & $\rho$\\
\hline
Canada	&	0.509	&	0.500	&	0.590	&	0.514	&	0.551	&	0.560	&	0.704	\\
France	&	0.396	&	0.462	&	0.441	&	0.425	&	0.413	&	0.413	&	0.543	\\
Germany	&	0.377	&	0.396	&	0.461	&	0.415	&	0.421	&	0.423	&	0.577	\\
UK	&	0.415 &	0.415	&	0.422	&	0.378	&	0.387	&	0.389	&	0.528	\\
Netherlands	& 0.340	&	0.415	&	0.431	&	0.401	&	0.373	&	0.380	&	0.535	\\
Sweden	&	0.340	&	0.368	&	0.380	&	0.342	&	0.369	&	0.367	&	0.493	\\
Belgium	&	0.340	&	0.434	&	0.344	&	0.344	&	0.350	&	0.351	&	0.470	\\
Switzerland	& 0.377	&	0.368	&	0.326	&	0.366	&	0.355	&	0.351	&	0.458	\\
Spain	&	0.377	&	0.358	&	0.368	&	0.353	&	0.336	&	0.336	&	0.494	\\
Norway	&	0.396	&	0.396	&	0.278	&	0.322	&	0.358	&	0.365	&	0.427	\\
Austria	&	0.396	&	0.377	&	0.257	&	0.329	&	0.341	&	0.339	&	0.401	\\
Italy	&	0.340	&	0.321	&	0.370	&	0.310	&	0.327	&	0.324	&	0.499	\\
Denmark	& 0.283	&	0.358	&	0.226	&	0.339	&	0.308	&	0.313	&	0.375	\\
Australia &	0.264	&	0.245	&	0.072	&	0.232	&	0.223	&	0.236	&	0.256	\\
Singapore &	0.245	&	0.217	&	0.120	&	0.196	&	0.219	&	0.209	&	0.281	\\
Hong Kong &	0.226	&	0.198	&	0.057	&	0.150	&	0.151	&	0.155	&	0.204	\\
Japan	& 0.113	&	0.123	&	0.001	&	0.122	&	0.121	&	0.120	&	0.054	\\
\hline
\end{tabular}
\caption{Lower TDC: US vs other developed countries, depending on the TDC estimator. The correlation coefficient is $\rho$. }\label{tab:lowerUS}
\end{table}

\begin{table}[htbp]
\centering
\begin{tabular}{@{}l|llllll|l@{}}
\hline
Country     & (1)  & (2)  & (3)  & (4)  & (5)  & (6)  & $\rho$ \\
\hline
Canada	&	0.377	&	0.434	&	0.530	&	0.431	&	0.455	&	0.455	&	0.704	\\
Germany	&	0.321	&	0.406	&	0.445	&	0.427	&	0.418	&	0.427	&	0.577	\\
Netherlands	& 0.321	&	0.340	&	0.420	&	0.417	&	0.418	&	0.418	&	0.535	\\
France	&	0.283	&	0.340	&	0.426	&	0.317	&	0.404	&	0.410	&	0.543	\\
UK	&	0.302	&	0.321	&	0.405	&	0.313	&	0.332	&	0.353	&	0.528	\\
Sweden	&	0.302	&	0.330	&	0.382	&	0.291	&	0.335	&	0.330	&	0.493	\\
Spain	&	0.283	&	0.292	&	0.372	&	0.295	&	0.320	&	0.313	&	0.494	\\
Belgium	&	0.264	&	0.302	&	0.356	&	0.284	&	0.330	&	0.330	&	0.470	\\
Italy	&	0.264	&	0.264	&	0.384	&	0.256	&	0.300	&	0.302	&	0.458	\\
Switzerland	& 0.302	&	0.236	&	0.336	&	0.242	&	0.269	&	0.268	&	0.499	\\
Norway	&	0.226	&	0.274	&	0.302	&	0.274	&	0.264	&	0.268	&	0.427	\\
Denmark	& 0.226	&	0.236	&	0.267	&	0.214	&	0.251	&	0.253	&	0.375	\\
Austria	& 0.132	&	0.236	&	0.282	&	0.240	&	0.257	&	0.257	&	0.401	\\
Singapore	& 0.226	&	0.264	&	0.200	&	0.219	&	0.233	&	0.233	&	0.281	\\
Australia	& 0.189	&	0.245	&	0.164	&	0.227	&	0.216	&	0.216	&	0.256	\\
Hong Kong	& 0.151	&	0.151	&	0.158	&	0.158	&	0.165	&	0.163	&	0.204	\\
Japan	&	0.075	&	0.113	&	0.052	&	0.103	&	0.105	&	0.105	&	0.054	\\ \hline
\end{tabular}
\caption{Upper TDC: US vs other developed countries, depending on the TDC estimator. The correlation coefficient is $\rho$. }\label{tab:upperUS}
\end{table}

We observe a global coherence between the results of the six estimators. The strongest discrepancies among the estimators appear for estimator (3), which tends to underestimate the lower TDC with respect to the other estimators, when the TDC is low, and which also tends to overestimate the upper TDC, except for the four countries with the lowest TDC. The arbitrary estimator (1) tends to underestimate the upper TDC. 

The three Asian markets (Japan, Hong Kong, Singapore) exhibit the lowest lower and upper TDC. Of course, the time zone difference with the US can cause date shifts: if the US market is driving the global economy, the effect on the Asian markets is to be observed one day later. 

Even though the lower TDC is globally higher than the upper TDC, we see that a strong upper TDC is generally related to a strong lower TDC. This result suggests that one cannot benefit from a pairwise boom without having a risk of simultaneous crash.

\section{Discussion and conclusion}

We have given an expression of the MSE for the nonparametric TDC estimator. By minimizing this MSE in the case of a Clayton or a Gumbel copula, we have proposed a semiparametric method for estimating either the lower or the upper TDC. It is based on a plug-in approach, in which the parameter of the Clayton or Gumbel copula is estimated on a well-selected part of the observations. A simulation study shows that this kind of estimator offers reasonably good performance. It behaves in a similar way as the method relying on the plateau-finding algorithm does, for data generated by various types of copulas. 

It is worth noting that a recent independent work also aims to find a theoretically justified threshold for the nonparametric TDC estimator~\citep{GKO}. Using theoretical results by~\cite{Segers} and assuming another regularity condition of the copula than in the present work, namely $\partial^2 C(u_j,u_k)/\partial u_j\partial u_k=\mathcal O(1/\sqrt{u_ju_k})$, this article proposes an approximation of the empirical copula process. The residual is bounded by a quantity going to zero when the number of observations tends to infinity. The asymptotic bias and variance of the TDC estimator are thus obtained by neglecting this residual term and by focusing on the approximation of the empirical copula process. The authors also evaluate several practical parameter selection methods: a fully nonparametric method, mixing a statistical bootstrap and a box kernel smoothing, and a parametric approach, specifically for the bias, without censoring. Beyond the theoretical and the methodological differences with our work, this article shows simulation results which are also close to the plateau-finding algorithm.

Our own simulation results, along with the relative simplicity of our method compared to the plateau-finding algorithm, which embeds additional smoothing into the basic TDC estimator, thus legitimize our new estimator. Therefore we recommend our approach when facing an unknown type of underlying distribution.

\section*{Acknowledgement}

The authors are grateful to Benjamin Bobbia and anonymous readers for valuable discussions and comments.

\subsection*{Availability of supporting data and code}

The generating engine used for the simulated datasets used in the current study is available in \url{https://github.com/maximenc/pycop}.

The empirical data that support the findings of this study are available from MSCI (\url{https://www.msci.com/our-solutions/indexes/developed-markets/}) with the software Macrobond (\url{https://www.macrobond.com/}) but restrictions apply to the availability of these data, which were used under licence for the current study, and so are not publicly available.

\bibliographystyle{plainnat} 
\bibliography{biblio}

\appendix

\section{Proof of Theorem~\ref{th:ErrorEstim}}\label{sec:proofErrorEstim}

\begin{proof}
We define a dependence parameter for a given probability $q$ as:
\begin{equation}\label{eq:lambdaU}
\lambda_L(q)=\frac{\delta(q)}{q}.
\end{equation}
In particular, $\lambda_L=\underset{q\rightarrow 0}{\lim} \lambda_L(q)$. Since $\delta(0)=0$ by a basic property of copulas, we also have $\lambda_L=\delta'(0)$. We decompose the error of the estimator provided in equation~\eqref{eq:NonParmLambdaL} in noise and bias:
$$\widehat \lambda_{L,n}\left(\frac{i(n)}{n}\right)-\lambda_L = \left(\widehat \lambda_{L,n}\left(\frac{i(n)}{n}\right)-\lambda_L\left(\frac{i(n)}{n}\right)\right) + \left(\lambda_L\left(\frac{i(n)}{n}\right)-\lambda_L\right).$$
Following the work of Fermanian \textit{et al.} and in particular their Theorem 3, we know that the empirical copula process $\sqrt{n}(\widehat C_n(u,v) -C(u,v))$ converges weakly towards a Gaussian process $G_C(u,v)$~\citep{FRW}:
$$G_C(u,v)=B_C(u,v)-h_1(u,v)B_C(u,1)-h_2(u,v)B_C(1,v),$$
where $B_C$ is a Brownian bridge on $[0,1]^2$ of covariance
\begin{equation}\label{eq:CovBrownBridge}
\E\left[B_C(u,v)B_C(u',v')\right]=C(u\wedge u',v\wedge v')-C(u,v)C(u',v').
\end{equation}
Therefore, $\sqrt{n}\left(\widehat \lambda_{L,n}\left(i(n)/n\right)-\lambda_L\left(i(n)/n\right)\right)=\sqrt{n}(\widehat C_n(\alpha,\alpha) -C(\alpha,\alpha))/\alpha$, weakly converges toward $G_C(\alpha,\alpha)/\alpha$. As a consequence, $n\left(\widehat \lambda_{L,n}\left(i(n)/n\right)-\lambda_L\left(i(n)/n\right)\right)^2$ weakly converges toward $G_C(\alpha,\alpha)^2/\alpha^2$~\citep[Th. 5.2]{Billingsley} and, thanks to the uniform integrability proved in Appendix~\ref{sec:unifint}, $\E\left[n\left(\widehat \lambda_{L,n}\left(i(n)/n\right)-\lambda_L\left(i(n)/n\right)\right)^2\right]$ converges toward $\E\left[G_C(\alpha,\alpha)^2\right]/\alpha^2$~\citep[Th. 5.4]{Billingsley}. We note that, for $u\in(0,1)$,
$$\begin{array}{ccl}
\E[G_C(u,u)^2] & = & \E[B_C(u,u)^2] + h_1(u)^2\E[B_C(u,1)^2] + h_2(u)^2\E[B_C(1,u)^2] \\
 & & - 2 h_1(u)\E[B_C(u,u)B_C(u,1)] - 2 h_2(u)\E[B_C(u,u)B_C(1,u)] \\
 & & + 2 h_1(u)h_2(u)\E[B_C(u,1)B_C(1,u)] \\
 & = & \delta(u)-\delta(u)^2 + h_1(u)^2\left(C(u,1)-C(u,1)^2\right) + h_2(u)^2\left(C(1,u)-C(1,u)^2\right) \\
 & & - 2 h_1(u)\left(\delta(u)-\delta(u)C(u,1)\right) - 2 h_2(u)\left(\delta(u)-\delta(u)C(1,u)\right) \\
 & & + 2 h_1(u)h_2(u)\left(\delta(u)-C(u,1)C(1,u)\right) \\
  & = & \delta(u)(1-\delta(u)) + h_1(u)^2u(1-u) + h_2(u)^2u(1-u) \\
 & & - 2 h_1(u)\delta(u)(1-u) - 2 h_2(u)\delta(u)(1-u) \\
 & & + 2 h_1(u)h_2(u)\left(\delta(u)-u^2\right) \\
 & = & \sigma^2(u),
\end{array}$$
according to equation~\eqref{eq:CovBrownBridge} and using the fact that $C(u,1)=C(1,u)=u$. As a consequence, the asymptotic MSE is
$$\frac{1}{n\alpha^2}\E\left[G_C\left(\alpha,\alpha\right)^2\right] +\left(\lambda_L\left(\alpha\right)-\lambda_L\right)^2 = \frac{1}{n\alpha^2}\sigma^2\left(\alpha\right) +\left(\frac{1}{\alpha}\delta\left(\alpha\right)-\delta'(0)\right)^2.$$
\end{proof}

\section{Uniform integrability of the empirical copula process}\label{sec:unifint}

In the proof of Theorem~\ref{th:ErrorEstim}, we have supposed that $n\left(\widehat{\delta}_n(\alpha) - \delta(\alpha)\right)^2$ is uniformly integrable. We now prove this property. It is in fact enough to show that \citep[page 32]{Billingsley}:
\begin{equation}\label{eq:unifintsuff}\exists\varepsilon>0,\ \sup_n \E\left(\left|n\left(\widehat{\delta}_n(\alpha) - \delta(\alpha)\right)^2\right|^{1+\varepsilon}\right)<\infty.
\end{equation}
More generally, let's consider $u,v\in[0,1]$ and the following decomposition~\citep[proof of Proposition 3.1]{Segers}:
\begin{equation}\label{eq:Segers}
\sqrt{n}\left(\widehat C_n(u,v)-C(u,v)\right)=\sqrt{n}\left(C_n(\hat u,\hat v)-C(\hat u,\hat v)\right)+\sqrt{n}\left(C(\hat u,\hat v)-C(u,v)\right),
\end{equation}
where $C_n(u,v)=\frac{1}{n}\sum_{j=1}^n{\indic\{F_{X}(X_j)\leq u,F_{Y}(Y_j)\leq v\}}$, $\hat u=F_X\left(\widehat F_{X,n}^{-1}(u)\right)$, and $\hat v=F_Y\left(\widehat F_{Y,n}^{-1}(v)\right)$. We're going to show that each of the two terms in equation~\eqref{eq:Segers} has a fourth moment uniformly bounded in $n$.

Denote by $\mathcal C_n$ the usual empirical copula process $\sqrt{n}(C_n-C)$. The first term in equation~\eqref{eq:Segers} is then $\mathcal C_n(\hat u,\hat v)$. It satisfies~\citep[Theorem 2.14.1]{VW}:
\begin{equation}\label{eq:Segers3}
\sup_n \E\left(|\mathcal C_n(\hat u,\hat v)|^4\right)\leq \sup_n \E\left(\sup_{u,v}|\mathcal C_n(u,v)|^4\right) < \infty.
\end{equation}

Regarding the second term in equation~\eqref{eq:Segers}, by Lipschitz continuity of copulas~\citep{SW,DNO}, we have
\begin{equation}\label{eq:Segers2}
\sqrt{n}\left|C(\hat u,\hat v)-C(u,v)\right|\leq \sqrt{n}\left|\hat u-u\right|+\sqrt{n}\left|\hat v-v\right|.
\end{equation}
There exists $i\in\llbracket 1,n\rrbracket$ such that $(i-1)/n<u\leq i/n$ and for which $\hat u=F_X(X_{(i)})$, where $X_{(i)}$ is the $i$-th order statistic. The variable $\hat u$ is itself an order statistic from a uniform distribution and it thus follows a Beta distribution of parameters $i$ and $n+1-i$~\citep{DN}. After some calculations using the moments of a Beta distribution, we find that the fourth centered moment $\E[(\hat u-i/(n+1))^4]$ is bounded up to a constant by $1/n^2$. Given that $|i/(n+1)-u|\leq 1/n$, we deduce that
$$\sup_n \E\left[\left(\sqrt{n}\left|\hat u-u\right|\right)^4\right] < \infty.$$
We obtain a similar result replacing $\hat u-u$ by $\hat v-v$, so that, using equation~\eqref{eq:Segers2}, we obtain
\begin{equation}\label{eq:Segers4}
\sup_n \E\left[\sqrt{n}\left|C(\hat u,\hat v)-C(u,v)\right|\right]<\infty.
\end{equation}
Equations~\eqref{eq:Segers3} and~\eqref{eq:Segers4} thus confirm the uniform integrability of the empirical copula process.

\section{Proof of Proposition~\ref{pro:Clayton}}\label{sec:proofClayton}

\begin{proof}
The diagonal section of the Clayton copula is obtained by considering the case $u=v$:
$$\delta(u)=\left(2u^{-\theta}-1\right)^{-1/\theta}.$$
Its first derivative is:
$$\delta'(u)=2u^{-\theta-1}\left(2u^{-\theta}-1\right)^{-1-1/\theta}=2\left(2-u^{\theta}\right)^{-1-1/\theta},$$
whose value in $u=0$ is $\delta'(0)=2^{-1/\theta}$. According to Theorem~\ref{th:ErrorEstim}, the asymptotic bias is thus:
$$\frac{1}{\alpha}\delta\left(\alpha\right)-\delta'(0)=\frac{1}{\alpha}\left(2\alpha^{-\theta}-1\right)^{-1/\theta}-2^{-1/\theta}=\left(2-\alpha^{\theta}\right)^{-1/\theta}-2^{-1/\theta}.$$
The corresponding h-function is~\citep{SS}:
$$h(u)=\frac{\delta'(u)}{2}=\delta(u)\frac{u^{-\theta-1}}{(2u^{-\theta}-1)}=\frac{\delta(u)}{u}\frac{1}{2-u^{\theta}}.$$
Following equation~\eqref{eq:SymVar}, we get:
$$\begin{array}{ccl}
\sigma^2(u) & = & \delta(u)(1-\delta(u)) +2(1-u)h(u)\left[uh(u)- 2 \delta(u)\right] + 2 h(u)^2(\delta(u)-u^2) \\
 & = &  \delta(u)(1-\delta(u)) +2(1-u)\frac{\delta(u)}{u}\frac{1}{2-u^{\theta}}\left[u\frac{\delta(u)}{u}\frac{1}{2-u^{\theta}}-2 \delta(u)\right] + 2 \left(\frac{\delta(u)}{u}\frac{1}{2-u^{\theta}}\right)^2(\delta(u)-u^2) \\
 & = &  \delta(u)(1-\delta(u)) +2\left(\frac{1}{u}-1\right)\delta(u)^2\frac{1}{2-u^{\theta}}\left[\frac{1}{2-u^{\theta}}-2 \right] + 2 \frac{\delta(u)^2}{(2-u^{\theta})^2}\left(\frac{\delta(u)}{u^2}-1\right)  \\
 & = &  \delta(u) + \delta(u)^2\left[-1+2\left(\frac{1}{u}-1\right)\frac{1}{2-u^{\theta}}\left(\frac{1}{2-u^{\theta}}-2 \right)-\frac{2}{(2-u^{\theta})^2}\right] + 2 \delta(u)^3\left[\frac{1}{u^2(2-u^{\theta})^2}\right] \\
 & = &  \delta(u) - \delta(u)^2\left[1+2\frac{2(1-u)(1-u^{\theta})+1}{u(2-u^{\theta})^2}\right] + 2\delta(u)^3\left[\frac{1}{u^2(2-u^{\theta})^2}\right].
\end{array}$$
\end{proof}

\section{Proof of Theorem~\ref{th:ErrorEstimU}}\label{sec:proofErrorEstimU}

\begin{proof}
We first focus on the variance part of the tail dependence estimator:
$$\E\left[\left(\widehat{\lambda}_{U,n}(i(n) / n)-\lambda_{U}(i(n) / n)\right)^2\right].$$ 
We know that $\sqrt{n}(\widehat C(u,v) -C(u,v))$ converges weakly towards a Gaussian process $G_C(u,v)$, as already mentioned in the proof of Theorem~\ref{th:ErrorEstim}. Therefore, following the same reasoning as in the proof of Theorem~\ref{th:ErrorEstim}, the second moment of $\widehat{\lambda}_{U}(i(n) / n)-\lambda_{U}(i(n) / n)$ converges toward the second moment of $G_{C}\left(\alpha,\alpha\right)\sqrt{n}/(n-\alpha n)$, which is of mean $0$ and of variance 
$$\frac{1}{1(1-\alpha)^{2}} \mathbb{E}\left[G_{C}\left(\alpha,\alpha\right)^{2}\right] =\frac{1}{n(1-\alpha)^{2}} \sigma^{2}\left(\alpha\right).$$

By noting that $\lambda_U=2-\delta'(1)$, the expression of the bias is straightforward and we conclude by noting that the MSE is the sum of the variance and of the squared bias.
\end{proof}

\section{Proof of Proposition~\ref{pro:Gumbel}}\label{sec:proofGumbel}

\begin{proof}
In the Gumbel case,
$$\delta(u)=C(u, u)= \exp \left[-\left\{ 2\left(-\ln \left(u\right)\right)^{\theta}\right\}^{\frac{1}{\theta}}\right] =
 \exp \left[- (2 t)^{\frac{1}{\theta}}\right],$$
where $t=\left(-\ln \left(u\right)\right)^{\theta}$. Its first derivative is $\delta'(u)=\frac{2^{1/\theta}\delta(u)}{u}$, and in particular $\delta'(1)=2^{1/\theta}$. According to Theorem~\ref{th:ErrorEstimU}, the asymptotic bias is thus:
$$ \frac{1-2 \alpha+\delta\left(\alpha\right)}{1-\alpha} - 2 + \delta'(1)= \frac{1-2 \alpha+\delta\left(\alpha\right)}{1-\alpha}-2 +2^{1/\theta}.$$
The h-function is~\citep{SS}:
$$h\left(u\right)=-\frac{\mathrm{e}^{-\left(2t\right)^{\frac{1}{\theta}}}\left(2t\right)^{\frac{1}{\theta}-1} t}{u \ln \left(u\right)} = \frac{\delta(u)}{u}2^{\frac{1}{\theta}-1},
$$
so that
$$\begin{array}{ccl}
\sigma^{2}(u) & = & \delta(u)(1-\delta(u))+2(1-u) h(u)[u h(u)-2\delta(u)]+2h(u)^{2}\left(\delta(u)-u^{2}\right) \\
 & = & \delta(u)(1-\delta(u)) + \delta(u)^2\left(\frac{1}{u}-1\right) 2^{\frac{1}{\theta}}\left(2^{\frac{1}{\theta}-1}-2\right) + \delta(u)^2 2^{\frac{2}{\theta}-1}\left(\frac{\delta(u)}{u^2}-1\right).
\end{array}$$
\end{proof}

\section{Proof of Theorem~\ref{th:ErrorEstimAverage}}\label{sec:proofErrorEstimAverage}

\begin{proof}
Like in Theorems~\ref{th:ErrorEstim} and~\ref{th:ErrorEstimU}, we decompose the MSE in variance and squared bias:
$$\begin{array}{ccl}
\E\left[\left(\frac{1}{m}\sum_{k=1}^{m}{\widehat \lambda_L\left(\frac{i_k(n)}{n}\right)}-\lambda_L\right)^2\right] & = & \E\left[\frac{1}{m}\sum_{k,l=1}^{m}{\left(\widehat \lambda_L\left(\frac{i_k(n)}{n}\right)-\lambda_L\left(\frac{i_k(n)}{n}\right)\right) \left(\widehat \lambda_L\left(\frac{i_l(n)}{n}\right)-\lambda_L\left(\frac{i_l(n)}{n}\right)\right)}\right] \\
 & & + \left[\left(\frac{1}{m}\sum_{k=1}^{m}{\lambda_L\left(\alpha_k\right)}-\lambda_L\right)^2\right],
\end{array}$$
with $\lambda_L(u)$ defined by equation~\eqref{eq:lambdaU}. Like in the previous theorems, the bias part finds a straightforward expression using $\delta$ and $\delta'$:
$$\left(\frac{1}{m}\sum_{k=1}^{m}{\lambda_L\left(\alpha_k\right)}-\lambda_L\right)^2 = \left(\frac{1}{m}\sum_{k=1}^{m}{\frac{1}{\alpha_k}\delta\left(\alpha_k\right)}-\delta'(0)\right)^2.$$

We now focus on the variance part of the MSE. Like in the proof of Theorem~\ref{th:ErrorEstim}, we note that the empirical copula process $\sqrt{n}(\widehat C(u,v) -C(u,v))$ converges weakly towards a Gaussian process $G_C(u,v)$:
$$G_C(u,v)=B_C(u,v)-h_1(u,v)B_C(u,1)-h_2(u,v)B_C(1,v),$$
where $B_C$ is a Brownian bridge on $[0,1]^2$ whose covariance is provided by equation~\eqref{eq:CovBrownBridge}. Therefore, the covariance between $G_C(u,u)$ and $G_C(v,v)$ is
$$\begin{array}{ccl}
\E[G_C(u,u)G_C(v,v)] & = & \E[B_C(u,u)B_C(v,v)] + h_1(u)h_1(v)\E[B_C(u,1)B_C(v,1)] + h_2(u)h_2(v)\E[B_C(1,u)B_C(1,v)] \\
 & & - h_1(v)\E[B_C(u,u)B_C(v,1)] - h_2(v)\E[B_C(u,u)B_C(1,v)] \\
 & & - h_1(u)\E[B_C(v,v)B_C(u,1)] - h_2(u)\E[B_C(v,v)B_C(1,u)] \\
 & & + h_1(u)h_2(v)\E[B_C(u,1)B_C(1,v)]+ h_1(v)h_2(u)\E[B_C(v,1)B_C(1,u)] \\
 & = & \delta(u\wedge v)-\delta(u)\delta(v) + (h_1(u)h_1(v)+h_2(u)h_2(v))((u\wedge v)-uv) \\
 & & - h_1(v)(C(u\wedge v,u)-v\delta(u)) - h_2(v)(C(u,u\wedge v)-v\delta(u)) \\
 & & - h_1(u)(C(u\wedge v,v)-u\delta(v)) - h_2(u)(C(v,u\wedge v)-u\delta(v)) \\
 & & + h_1(u)h_2(v)(C(u,v)-uv)+ h_1(v)h_2(u)(C(v,u)-uv)) \\
 & = & \mathcal K(u,v),
\end{array}$$
according to equation~\eqref{eq:CovBrownBridge} and using the fact that $C(u,1)=C(1,u)=u$. Following the same reasoning as in the proof of Theorem~\ref{th:ErrorEstim}, this result leads to the expression of the asymptotic variance of the TDC estimator provided in Theorem~\ref{th:ErrorEstimAverage} thanks to equation~\eqref{eq:lambdaU}, which links $(\widehat{\lambda}_L(u)-\lambda_L(u))$ to $(\widehat{\delta}(u)-\delta(u))/u$, that is to $(\widehat C(u,u)-C(u,u))/u$.
\end{proof}

\section{Tables of results for the simulation study}

\begin{table}[htbp]
\resizebox{\textwidth}{!}{%
\begin{tabular}{lcccccccccccc}
\hline
\multirow{2}{*}{Dataset} & \multicolumn{1}{c}{\multirow{2}{*}{Method}} & \multicolumn{3}{c}{$n = 500$} & \multicolumn{3}{c}{$n=1000$}  & \multicolumn{3}{c}{$n=2000$}  \\
                         & \multicolumn{1}{c}{}                        & $\text{Bias}$ & $\sigma(\widehat{\lambda}_{U,n})$ & $\text{RMSE}$ & $\text{Bias}$ &  $\sigma(\widehat{\lambda}_{U,n})$ & $\text{RMSE}$ & $\text{Bias}$ &  $\sigma(\widehat{\lambda}_{U,n})$ & $\text{RMSE}$ \\ \hline
 $\theta=1.10$ &    (1) &     0.08 &      0.19 &     0.21 &      0.09 &       0.11 &      0.14 &      0.09 &       0.09 &      0.12 \\
$\lambda_U=0.12$ &    (2) &     0.11 &      0.12 &     0.17 &      0.09 &       0.08 &      0.12 &      0.12 &       0.06 &      0.13 \\
           &    (3) &     0.15 &      0.03 &     0.15 &      0.15 &       0.03 &      0.15 &      0.15 &       0.02 &      0.15 \\
           &    (4) &     0.18 &      0.06 &     0.19 &      0.17 &       0.06 &      0.18 &      0.16 &       0.06 &      0.17 \\
           &    (5) &     0.18 &      0.07 &     0.20 &      0.17 &       0.06 &      0.18 &      0.16 &       0.04 &      0.16 \\
           &    (6) &     0.18 &      0.07 &     0.20 &      0.17 &       0.06 &      0.18 &      0.15 &       0.04 &      0.16 \\  \hline
 $\theta=1.50$ &    (1) &    -0.24 &      0.17 &     0.29 &     -0.22 &       0.12 &      0.25 &     -0.22 &       0.08 &      0.23 \\
$\lambda_U=0.41$&    (2) &    -0.19 &      0.12 &     0.23 &     -0.20 &       0.08 &      0.22 &     -0.19 &       0.07 &      0.20 \\
           &    (3) &    -0.10 &      0.03 &     0.10 &     -0.09 &       0.02 &      0.10 &     -0.09 &       0.02 &      0.10 \\
           &    (4) &    -0.12 &      0.06 &     0.13 &     -0.13 &       0.06 &      0.14 &     -0.14 &       0.05 &      0.15 \\
           &    (5) &    -0.11 &      0.07 &     0.13 &     -0.13 &       0.06 &      0.14 &     -0.14 &       0.04 &      0.15 \\
           &    (6) &    -0.11 &      0.06 &     0.13 &     -0.13 &       0.06 &      0.14 &     -0.14 &       0.04 &      0.15 \\  \hline
$\theta=1.75$ &    (1) &    -0.32 &      0.17 &     0.36 &     -0.29 &       0.13 &      0.31 &     -0.29 &       0.08 &      0.30 \\
  $\lambda_U=0.51$ &    (2) &    -0.25 &      0.14 &     0.29 &     -0.25 &       0.09 &      0.26 &     -0.26 &       0.06 &      0.26 \\
           &    (3) &    -0.12 &      0.03 &     0.12 &     -0.12 &       0.02 &      0.12 &     -0.12 &       0.02 &      0.12 \\
           &    (4) &    -0.16 &      0.08 &     0.18 &     -0.18 &       0.07 &      0.19 &     -0.18 &       0.06 &      0.19 \\
           &    (5) &    -0.14 &      0.07 &     0.16 &     -0.16 &       0.05 &      0.17 &     -0.18 &       0.04 &      0.19 \\
           &    (6) &    -0.14 &      0.06 &     0.16 &     -0.16 &       0.05 &      0.17 &     -0.18 &       0.04 &      0.18 \\  \hline
   $\theta=2.00$ &    (1) &    -0.28 &      0.20 &     0.34 &     -0.28 &       0.12 &      0.30 &     -0.28 &       0.08 &      0.30 \\
 $\lambda_U=0.59$ &    (2) &    -0.24 &      0.14 &     0.28 &     -0.25 &       0.10 &      0.27 &     -0.25 &       0.07 &      0.26 \\
           &    (3) &    -0.11 &      0.02 &     0.11 &     -0.11 &       0.02 &      0.11 &     -0.11 &       0.01 &      0.11 \\
           &    (4) &    -0.14 &      0.06 &     0.16 &     -0.17 &       0.06 &      0.18 &     -0.18 &       0.06 &      0.19 \\
           &    (5) &    -0.13 &      0.07 &     0.15 &     -0.15 &       0.05 &      0.16 &     -0.17 &       0.04 &      0.17 \\
           &    (6) &    -0.12 &      0.06 &     0.14 &     -0.15 &       0.05 &      0.16 &     -0.17 &       0.04 &      0.17 \\
\hline
\end{tabular}}
\caption{Upper tail dependence with 100 Gumbel simulations. }\label{tab:gumb}
\end{table}

\begin{table}[htbp]
\resizebox{\textwidth}{!}{%
\begin{tabular}{lcccccccccccc}
\hline
\multirow{2}{*}{Dataset} & \multicolumn{1}{c}{\multirow{2}{*}{Method}} & \multicolumn{3}{c}{$n = 500$} & \multicolumn{3}{c}{$n=1000$}  & \multicolumn{3}{c}{$n=2000$}  \\
                         & \multicolumn{1}{c}{}                        & $\text{Bias}$ & $\sigma(\widehat{\lambda}_{U,n})$ & $\text{RMSE}$ & $\text{Bias}$ &  $\sigma(\widehat{\lambda}_{U,n})$ & $\text{RMSE}$ & $\text{Bias}$ &  $\sigma(\widehat{\lambda}_{U,n})$ & $\text{RMSE}$ \\ \hline
   $\rho=0$ &    (1) &     0.00 &      0.19 &     0.19 &     -0.01 &       0.13 &      0.13 &     -0.01 &       0.10 &      0.10 \\
    $\nu=1$ &    (2) &    -0.01 &      0.13 &     0.13 &      0.01 &       0.08 &      0.09 &     -0.01 &       0.07 &      0.07 \\
$\lambda_U=0.29$  &    (3) &    -0.19 &      0.04 &     0.19 &     -0.19 &       0.03 &      0.19 &     -0.19 &       0.02 &      0.19 \\
         &    (4) &    -0.02 &      0.07 &     0.07 &     -0.00 &       0.05 &      0.05 &     -0.01 &       0.05 &      0.05 \\
         &    (5) &    -0.01 &      0.09 &     0.09 &      0.01 &       0.06 &      0.06 &     -0.00 &       0.06 &      0.06 \\
         &    (6) &    -0.01 &      0.08 &     0.09 &      0.01 &       0.06 &      0.06 &     -0.00 &       0.06 &      0.06 \\ \hline
   $\rho=0$ &    (1) &    -0.04 &      0.14 &     0.14 &      0.01 &       0.11 &      0.11 &     -0.02 &       0.08 &      0.08 \\
    $\nu=2$ &    (2) &    -0.02 &      0.10 &     0.10 &      0.00 &       0.08 &      0.08 &     -0.00 &       0.06 &      0.06 \\
$\lambda_U=0.18$ &    (3) &    -0.11 &      0.04 &     0.12 &     -0.11 &       0.02 &      0.12 &     -0.11 &       0.02 &      0.11 \\
         &    (4) &     0.00 &      0.07 &     0.07 &      0.00 &       0.05 &      0.05 &      0.01 &       0.05 &      0.05 \\
         &    (5) &    -0.00 &      0.09 &     0.09 &      0.00 &       0.07 &      0.07 &      0.01 &       0.06 &      0.06 \\
         &    (6) &    -0.00 &      0.09 &     0.09 &      0.00 &       0.07 &      0.07 &      0.01 &       0.06 &      0.06 \\ \hline
   $\rho=0$ &    (1) &    -0.01 &      0.13 &     0.13 &      0.01 &       0.11 &      0.11 &      0.00 &       0.07 &      0.07 \\
    $\nu=3$ &    (2) &     0.00 &      0.10 &     0.10 &      0.01 &       0.07 &      0.07 &      0.01 &       0.05 &      0.05 \\ 
$\lambda_U=0.12$  &    (3) &    -0.06 &      0.04 &     0.07 &     -0.06 &       0.02 &      0.06 &     -0.06 &       0.02 &      0.07 \\
         &    (4) &     0.03 &      0.07 &     0.08 &      0.04 &       0.05 &      0.06 &      0.02 &       0.04 &      0.04 \\
         &    (5) &     0.01 &      0.08 &     0.09 &      0.02 &       0.06 &      0.07 &      0.02 &       0.05 &      0.06 \\
         &    (6) &     0.01 &      0.09 &     0.09 &      0.02 &       0.06 &      0.07 &      0.02 &       0.05 &      0.05 \\
\hline
\end{tabular}}
\caption{Upper tail dependence with 100 Student simulations, with $\rho=0$.}\label{tab:student}
\end{table}

\begin{table}[htbp]
\resizebox{\textwidth}{!}{%
\begin{tabular}{lcccccccccccc}
\hline
\multirow{2}{*}{Dataset} & \multicolumn{1}{c}{\multirow{2}{*}{Method}} & \multicolumn{3}{c}{$n = 500$} & \multicolumn{3}{c}{$n=1000$}  & \multicolumn{3}{c}{$n=2000$}  \\
                         & \multicolumn{1}{c}{}                        & $\text{Bias}$ & $\sigma(\widehat{\lambda}_{U,n})$ & $\text{RMSE}$ & $\text{Bias}$ &  $\sigma(\widehat{\lambda}_{U,n})$ & $\text{RMSE}$ & $\text{Bias}$ &  $\sigma(\widehat{\lambda}_{U,n})$ & $\text{RMSE}$ \\ \hline
$\rho=0.25$ &    (1) &    -0.01 &      0.20 &     0.20 &     -0.01 &       0.13 &      0.13 &     -0.02 &       0.10 &      0.11 \\
    $\nu=1$ &    (2) &    -0.02 &      0.14 &     0.14 &      0.01 &       0.09 &      0.09 &     -0.01 &       0.07 &      0.07 \\
$\lambda_U=0.39$  &    (3) &    -0.12 &      0.04 &     0.13 &     -0.12 &       0.03 &      0.13 &     -0.12 &       0.02 &      0.12 \\
         &    (4) &    -0.02 &      0.07 &     0.07 &     -0.01 &       0.05 &      0.05 &     -0.01 &       0.05 &      0.05 \\
         &    (5) &    -0.00 &      0.07 &     0.07 &      0.00 &       0.05 &      0.05 &     -0.00 &       0.04 &      0.04 \\
         &    (6) &    -0.00 &      0.07 &     0.07 &      0.00 &       0.05 &      0.05 &     -0.00 &       0.04 &      0.04 \\ \hline
$\rho=0.25$ &    (1) &    -0.03 &      0.16 &     0.17 &     -0.00 &       0.12 &      0.12 &     -0.00 &       0.09 &      0.09 \\
    $\nu=2$ &    (2) &    -0.03 &      0.11 &     0.11 &     -0.00 &       0.10 &      0.10 &      0.01 &       0.07 &      0.07 \\
$\lambda_U=0.27$ &    (3) &    -0.04 &      0.04 &     0.06 &     -0.04 &       0.03 &      0.05 &     -0.04 &       0.02 &      0.04 \\
         &    (4) &     0.01 &      0.06 &     0.06 &      0.01 &       0.05 &      0.05 &      0.01 &       0.05 &      0.05 \\
         &    (5) &     0.02 &      0.08 &     0.08 &      0.01 &       0.05 &      0.05 &      0.02 &       0.05 &      0.05 \\
         &    (6) &     0.02 &      0.07 &     0.08 &      0.01 &       0.05 &      0.05 &      0.02 &       0.05 &      0.05 \\ \hline
$\rho=0.25$ &    (1) &    -0.00 &      0.16 &     0.16 &      0.01 &       0.13 &      0.13 &     -0.00 &       0.08 &      0.08 \\
    $\nu=3$ &    (2) &     0.00 &      0.12 &     0.12 &      0.03 &       0.09 &      0.10 &      0.02 &       0.06 &      0.06 \\
$\lambda_U=0.20$ &    (3) &     0.02 &      0.04 &     0.05 &      0.03 &       0.03 &      0.04 &      0.02 &       0.02 &      0.03 \\
         &    (4) &     0.04 &      0.08 &     0.09 &      0.05 &       0.06 &      0.07 &      0.04 &       0.05 &      0.06 \\
         &    (5) &     0.06 &      0.09 &     0.10 &      0.06 &       0.05 &      0.08 &      0.04 &       0.04 &      0.06 \\
         &    (6) &     0.05 &      0.08 &     0.10 &      0.06 &       0.05 &      0.07 &      0.04 &       0.04 &      0.06 \\
\hline
\end{tabular}}
\caption{Upper tail dependence with 100 Student simulations, with $\rho=0.25$.}\label{tab:studentRho}
\end{table}

\begin{table}[htbp]
\resizebox{\textwidth}{!}{%
\begin{tabular}{lcccccccccccc}
\hline
\multirow{2}{*}{Dataset} & \multicolumn{1}{c}{\multirow{2}{*}{Method}} & \multicolumn{3}{c}{$n = 500$} & \multicolumn{3}{c}{$n=1000$}  & \multicolumn{3}{c}{$n=2000$}  \\
                         & \multicolumn{1}{c}{}                        & $\text{Bias}$ & $\sigma(\widehat{\lambda}_{U,n})$ & $\text{RMSE}$ & $\text{Bias}$ &  $\sigma(\widehat{\lambda}_{U,n})$ & $\text{RMSE}$ & $\text{Bias}$ &  $\sigma(\widehat{\lambda}_{U,n})$ & $\text{RMSE}$ \\ \hline
$\theta=0.1$ &    (1) &     0.04 &      0.08 &     0.09 &      0.06 &       0.07 &      0.09 &      0.04 &       0.04 &      0.06 \\
$\lambda_U=0.00$  &    (2) &     0.05 &      0.07 &     0.08 &      0.07 &       0.06 &      0.09 &      0.06 &       0.04 &      0.07 \\
          &    (3) &     0.06 &      0.04 &     0.07 &      0.06 &       0.02 &      0.07 &      0.06 &       0.02 &      0.06 \\
          &    (4) &     0.10 &      0.07 &     0.12 &      0.09 &       0.05 &      0.11 &      0.08 &       0.04 &      0.09 \\
          &    (5) &     0.08 &      0.07 &     0.11 &      0.08 &       0.05 &      0.10 &      0.07 &       0.03 &      0.07 \\
          &    (6) &     0.08 &      0.07 &     0.10 &      0.08 &       0.05 &      0.10 &      0.06 &       0.03 &      0.07 \\ \hline
$\theta=0.5$ &    (1) &     0.01 &      0.18 &     0.18 &      0.01 &       0.12 &      0.12 &      0.02 &       0.10 &      0.10 \\
$\lambda_U=0.25$ &    (2) &     0.04 &      0.14 &     0.15 &      0.04 &       0.09 &      0.10 &      0.03 &       0.05 &      0.06 \\
          &    (3) &     0.02 &      0.04 &     0.04 &      0.02 &       0.02 &      0.03 &      0.02 &       0.02 &      0.02 \\
          &    (4) &     0.08 &      0.07 &     0.10 &      0.06 &       0.06 &      0.08 &      0.05 &       0.05 &      0.07 \\
          &    (5) &     0.08 &      0.08 &     0.12 &      0.07 &       0.05 &      0.09 &      0.05 &       0.04 &      0.06 \\
          &    (6) &     0.09 &      0.08 &     0.12 &      0.07 &       0.05 &      0.09 &      0.05 &       0.04 &      0.06 \\ \hline
  $\theta=1.0$ &    (1) &    -0.00 &      0.17 &     0.17 &     -0.02 &       0.13 &      0.13 &     -0.02 &       0.10 &      0.11 \\
$\lambda_U=0.50$  &    (2) &    -0.00 &      0.13 &     0.13 &      0.01 &       0.10 &      0.10 &     -0.01 &       0.07 &      0.07 \\
          &    (3) &    -0.07 &      0.03 &     0.08 &     -0.08 &       0.05 &      0.09 &     -0.07 &       0.01 &      0.07 \\
          &    (4) &     0.00 &      0.05 &     0.05 &      0.00 &       0.05 &      0.05 &     -0.00 &       0.04 &      0.04 \\
          &    (5) &     0.03 &      0.05 &     0.06 &      0.02 &       0.05 &      0.05 &      0.01 &       0.04 &      0.05 \\
          &    (6) &     0.02 &      0.06 &     0.06 &      0.02 &       0.05 &      0.06 &      0.01 &       0.04 &      0.04 \\ \hline
$\theta=1.5$ &    (1) &    -0.02 &      0.17 &     0.17 &     -0.01 &       0.13 &      0.13 &      0.00 &       0.09 &      0.09 \\
$\lambda_U=0.63$  &    (2) &    -0.04 &      0.12 &     0.13 &     -0.02 &       0.09 &      0.09 &     -0.00 &       0.06 &      0.06 \\
          &    (3) &    -0.10 &      0.02 &     0.10 &     -0.10 &       0.02 &      0.10 &     -0.10 &       0.01 &      0.10 \\
          &    (4) &    -0.02 &      0.04 &     0.05 &     -0.02 &       0.04 &      0.04 &     -0.01 &       0.03 &      0.03 \\
          &    (5) &     0.01 &      0.05 &     0.05 &      0.00 &       0.04 &      0.04 &      0.00 &       0.03 &      0.03 \\
          &    (6) &     0.01 &      0.05 &     0.05 &      0.00 &       0.04 &      0.04 &      0.00 &       0.03 &      0.03 \\
\hline
\end{tabular}}
\caption{Upper tail dependence with 100 rotated Clayton simulations. }\label{tab:clayton}
\end{table}

\begin{table}[htbp]
\resizebox{\textwidth}{!}{%
\begin{tabular}{lcccccccccccc}
\hline
\multirow{2}{*}{Dataset} & \multicolumn{1}{c}{\multirow{2}{*}{Method}} & \multicolumn{3}{c}{$n = 500$} & \multicolumn{3}{c}{$n=1000$}  & \multicolumn{3}{c}{$n=2000$}  \\
                         & \multicolumn{1}{c}{}                        & $\text{Bias}$ & $\sigma(\widehat{\lambda}_{U,n})$ & $\text{RMSE}$ & $\text{Bias}$ &  $\sigma(\widehat{\lambda}_{U,n})$ & $\text{RMSE}$ & $\text{Bias}$ &  $\sigma(\widehat{\lambda}_{U,n})$ & $\text{RMSE}$ \\ \hline

$\rho=0.00$ &    (1) &     0.02 &      0.05 &     0.06 &      0.01 &       0.03 &      0.03 &      0.01 &       0.02 &      0.03 \\
         &    (2) &     0.02 &      0.04 &     0.05 &      0.02 &       0.03 &      0.04 &      0.02 &       0.02 &      0.03 \\
         &    (3) &     0.01 &      0.02 &     0.03 &      0.01 &       0.01 &      0.01 &      0.01 &       0.01 &      0.01 \\
         &    (4) &     0.04 &      0.05 &     0.06 &      0.03 &       0.04 &      0.05 &      0.02 &       0.02 &      0.03 \\
         &    (5) &     0.03 &      0.05 &     0.06 &      0.02 &       0.03 &      0.03 &      0.01 &       0.03 &      0.03 \\
         &    (6) &     0.02 &      0.05 &     0.05 &      0.02 &       0.03 &      0.04 &      0.01 &       0.02 &      0.02 \\  \hline
$\rho=0.25$ &    (1) &     0.05 &      0.10 &     0.11 &      0.05 &       0.06 &      0.08 &      0.05 &       0.04 &      0.06 \\
         &    (2) &     0.08 &      0.08 &     0.11 &      0.07 &       0.05 &      0.09 &      0.07 &       0.04 &      0.08 \\
         &    (3) &     0.18 &      0.04 &     0.19 &      0.18 &       0.02 &      0.18 &      0.18 &       0.02 &      0.18 \\
         &    (4) &     0.12 &      0.07 &     0.14 &      0.12 &       0.06 &      0.13 &      0.11 &       0.05 &      0.12 \\
         &    (5) &     0.14 &      0.07 &     0.15 &      0.12 &       0.05 &      0.13 &      0.10 &       0.03 &      0.10 \\
         &    (6) &     0.14 &      0.07 &     0.15 &      0.12 &       0.05 &      0.13 &      0.10 &       0.03 &      0.10 \\  \hline
 $\rho=0.50$  &    (1) &     0.12 &      0.14 &     0.19 &      0.12 &       0.09 &      0.15 &      0.14 &       0.07 &      0.15 \\
         &    (2) &     0.17 &      0.11 &     0.20 &      0.16 &       0.07 &      0.18 &      0.17 &       0.06 &      0.18 \\
         &    (3) &     0.38 &      0.03 &     0.38 &      0.38 &       0.02 &      0.38 &      0.37 &       0.02 &      0.37 \\
         &    (4) &     0.28 &      0.08 &     0.30 &      0.26 &       0.07 &      0.27 &      0.24 &       0.07 &      0.25 \\
         &    (5) &     0.30 &      0.07 &     0.31 &      0.28 &       0.06 &      0.28 &      0.25 &       0.04 &      0.25 \\
         &    (6) &     0.30 &      0.06 &     0.30 &      0.27 &       0.05 &      0.28 &      0.25 &       0.04 &      0.26 \\ \hline
$\rho=0.75$ &    (1) &     0.29 &      0.16 &     0.33 &      0.32 &       0.13 &      0.34 &      0.32 &       0.10 &      0.34 \\
         &    (2) &     0.35 &      0.13 &     0.38 &      0.36 &       0.09 &      0.37 &      0.36 &       0.07 &      0.37 \\
         &    (3) &     0.59 &      0.02 &     0.59 &      0.58 &       0.06 &      0.59 &      0.59 &       0.01 &      0.59 \\
         &    (4) &     0.48 &      0.06 &     0.49 &      0.47 &       0.08 &      0.48 &      0.46 &       0.06 &      0.46 \\
         &    (5) &     0.52 &      0.06 &     0.52 &      0.50 &       0.04 &      0.50 &      0.47 &       0.05 &      0.47 \\
         &    (6) &     0.53 &      0.05 &     0.53 &      0.50 &       0.04 &      0.51 &      0.48 &       0.04 &      0.48 \\
\hline
\end{tabular}}
\caption{Upper tail dependence with 100 Gaussian simulations.}\label{tab:gauss}
\end{table}

\end{document}